\documentclass
[superscriptaddress,secnumarabic,amssymb,amsmath,nobibnotes,aps,prd,showkeys,showpacs,nofootinbib,onecolumn]{revtex4}%
\usepackage{graphicx}
\usepackage{epsf}
\usepackage{bm}
\usepackage{amsmath}
\usepackage{amsfonts}
\usepackage{amssymb}%
\providecommand{\U}[1]{\protect\rule{.1in}{.1in}}

\newcommand{\be}{\begin{equation}}
\newcommand{\ee}{\end{equation}}

\newcommand{\mincir}{\raise
-3.truept\hbox{\rlap{\hbox{$\sim$}}\raise4.truept\hbox{$<$}\ }}
\newcommand{\magcir}{\raise
-3.truept\hbox{\rlap{\hbox{$\sim$}}\raise4.truept\hbox{$>$}\ }}

\begin{document}
\title{Cartan symmetries and global dynamical systems analysis in a higher-order
modified teleparallel theory}
\author{L. Karpathopoulos}
\affiliation{Faculty of Physics, Department of Astronomy-Astrophysics-Mechanics University
of Athens, Panepistemiopolis, Athens 157 83, Greece}
\author{S. Basilakos}
\email{svasil@academyofathens.gr}
\affiliation{Academy of Athens, Research Center for Astronomy and Applied Mathematics,
Soranou Efesiou 4, 11527, Athens, Greece}
\author{G. Leon}
\email{genly.leon@ucn.cl}
\affiliation{Departamento de Matem\'{a}ticas, Universidad Cat\'{o}lica del Norte, Avda.
Angamos 0610, Casilla 1280 Antofagasta, Chile}
\author{A. Paliathanasis}
\email{anpaliat@phys.uoa.gr}
\affiliation{Instituto de Ciencias F\'{\i}sicas y Matem\'{a}ticas, Universidad Austral de
Chile, Valdivia, Chile}
\affiliation{Department of Mathematics and Natural Sciences, Core Curriculum Program,
Prince Mohammad Bin Fahd University, Al Khobar 31952, Kingdom of Saudi Arabia}
\affiliation{Institute of Systems Science, Durban University of Technology, PO Box 1334,
Durban 4000, Republic of South Africa}
\author{M. Tsamparlis}
\email{mtsampa@phys.uoa.gr}
\affiliation{Faculty of Physics, Department of Astronomy-Astrophysics-Mechanics University
of Athens, Panepistemiopolis, Athens 157 83, Greece}
\keywords{Cosmology; Symmetries; Cartan symmetries; Teleparallel; Critical Points}
\pacs{98.80.-k, 95.35.+d, 95.36.+x}

\begin{abstract}
In a higher-order modified teleparallel theory cosmological we present
analytical cosmological solutions. In particular we determine forms of the
unknown potential which drives the scalar field such that the field equations
form a Liouville integrable system. For the determination of the conservation
laws we apply the Cartan symmetries. Furthermore, inspired from our solutions,
a toy model is studied and it is shown that it can describe the Supernova
data, while at the same time introduces dark matter components in the Hubble
function. When the extra matter source is a stiff fluid then we show how
analytical solutions for Bianchi I universes can be constructed from our
analysis. Finally, we perform a global dynamical analysis of the field
equations by using variables different from that of the Hubble-normalization.

\end{abstract}
\maketitle
\date{\today}

\section{Introduction}

A plethora of mechanisms has been introduced in order to explain the recent
cosmological observations \cite{Teg,Kowal,Komatsu,Ade15}. In particular, the
observed late-time acceleration of the universe has been attributed to a new
matter source which has been called dark energy. Several models have been
proposed for the dark energy among which scalar fields (quintessence, phantom
fields, k-essence), fluids with time-varying equation of state parameters
(Chaplygin gases) and the list goes on
\cite{Ratra,fere,Overduin,Linder,Lima,sahni,Kame,Gal1,Barrow,Brook,aasen,barea,lij,lij2,lij3,genly0,Leon:2009rc}%
.

However, there is a large body of dark energy models which have geometric
origin. In this class of scenarios the dark energy components correspond to
the new degrees of freedom in the field equations, which are introduced by the
modification of Einstein's General Relativity. For instance, the introduction
of quantum corrections in the Einstein's General Relativity is performed with
the use of higher-order invariants, such as polynomial terms involving the
Ricci scalar, the Gauss Bonnet term and many others. These considerations have
lead to the so called $f-$theories, in which a function $f\left(  X\right)  $
is introduced in the Einstein-Hilbert action where $X$ is a geometric
invariant
\cite{Buda,Sotiriou,odin1,Ferraro,mod1,on2,on3,mod2,mod3,nes3,Clifton,Clif2,Clif3,bcl,wang,olmo1,genly1,genly2}%
.

The Einstein-Hilbert action is not the unique action which provides the field
equations of General Relativity. The Palatini formalism \cite{palatini1} and
the teleparallel equivalent of GR (TEGR) \cite{ein28} are two alternative
variations which also under certain constraints lead again to General
Relativity (for more details see \cite{palatini2,maluf}). In TEGR the scalar
invariant $T$ of the Weitzenb\"{o}ck connection is considered as the
Lagrangian density of the field equations, while in the Palatini formalism the
metric and the connection are varied independently.

In this article, we work in the context of the TEGR by considering a
higher-order theory of gravity which introduces a scalar field with a
noncanonical kinetic term as a dark-energy component \cite{an1,anbas}%
\footnote{In the following, with the term \textquotedblleft
canonical\textquotedblright\ scalar field we refer to the quintessence scalar
field with a canonical kinetic term, that is, with Lagrangian $L_{\phi}=K-V$%
.}. Another well-known scalar field which is related with a modified theory is
the field in the O'Hanlon theory which describes the geometrodynamic degrees
of freedom in \thinspace\thinspace$f\left(  R\right)  -$gravity. More details
are given below.

We show that the field equations can be written with the use of a point-like
Lagrangian. Which describes the classical analogue for the motion of two
particles under an interaction between them, that is under the existence of an
effective potential. The kinetic term of the point-like Lagrangian describes
the degrees of freedom which concern the spacetime and the field, while the
effective potential is related to those which drive the evolution of the new
(noncanonical) scalar field.

We determine the special forms of this effective potential by requiring that
the field equations are Liouville integrable and the solution of the
Hamilton-Jacobi equation can be written in a closed-form expression. In order
the latter to be possible, constraints on the action have to be determined
which are equivalent to the existence of conservation laws. We derive these
conservation laws and prove the Liouville integrability of the corresponding
models by using the method of symmetries.

The symmetries, that is, the transformations which leave invariant a set of
differential equations is a powerful method for the determination of
conservation laws and exact solutions. The simplest type of symmetries are the
Lie point symmetries which have been applied in various problems for the
determination of new exact solutions either in the classical or at the quantum
level \cite{Maharaj1,Maharaj2,Maharaj3,lie1,lie2,lie3,lie5}. For the
gravitational models described by a point Lagrangian a special type of Lie
point symmetries are those which in addition leave invariant the action
integral. These Lie symmetries are called Noether symmetries which by means of
Noether's theorem allow the determination of conservation laws, for instance
see
\cite{Rosquist,Cap97M,nor3,nor2,Basilakos1,nor5,nor9,nor10,vak3,nor12,nor11,nor14,nor1,terzis2,nor7,nor6,nor8,gian,bah}%
. In the present work we choose to work with the Cartan's method in order to
study the symmetries of our field equations. The Cartan symmetries are based
on the invariance of the Cartan 1-form which is defined directly from the
Lagrangian under point transformations with generators in the tangent bundle
\cite{Crampin0,Crampin01,Crampin1,Sarlet}. It can be shown that for holonomic
dynamical systems the Cartan symmetries are equivalent to the so-called
generalized Noether symmetries \cite{marmo}, whereas for nonholonomic
dynamical systems the situation is different \cite{crampin22}.

Moreover, we perform a global dynamical analysis for that modified theory by
using a different set of variables from the Hubble normalization. We see that
the results of \cite{an1} are recovered; however new critical points are
derived while we show that there can be physical processes which were not
derived before. For instance, we show that it is possible the universe to pass
from an accelerate phase to a decelerate phase and vice verca. Furthermore, we
apply the results\ of the dynamical analysis to study the physical properties
of the theories which followed from the symmetry analysis while the field
equations are Liouville integrable. The plan of the paper is as follows.

In Section \ref{field}, we present the cosmological model of our study. The
field equations and the Lagrangian description in the minisuperspace approach
are derived for a spatially flat Friedmann-Lema\^{\i}tre-Robertson-Walker
(FLRW) universe with an ideal gas as an extra matter source. Furthermore, we
consider the Lagrangian for the case of Bianchi I models and show that the
field equations are reduced to that of FLRW when the ideal gas is a stiff
fluid. In Section \ref{cartan} we discuss briefly the Cartan formalism which
we use in the determination of the conservation laws. In Section
\ref{cartan2}, we determine the specific forms of the unknown parameters of
our model by requiring that the field equations admit Cartan symmetries.
Moreover, the method of Hamilton-Jacobi is applied in order to reduce the
field equations to a system of two first-order ordinary differential equations
and when it is feasible to write the closed-form solution of the system. The
dynamical system analysis is performed in Section \ref{SECT:II}.\ We see that
there is a specific potential for which the dimension of the dynamical system
is reduced. The critical points are derived for all the possible families of
theories as also their stability conditions. In Section \ref{cartan3} we
consider a closed-form solution from the previous section as a toy model and
we perform a likelihood analysis with the SNIa data. We find that the model
fits the SNIa data while at the same time the theory provides the dark matter
components in the Hubble function. In Appendix A we give the critical points
of the Hubble normalization and in Appendix B the critical points in the state
space of observable quantities. In the Section \ref{Sect:7}, we investigate
the evolution of the observables, the so called age parameter $\alpha=t H$,
the deceleration parameter $q$, and the fractional energy of scalar field and
Hubble-normalized kinetic term in a phase space. Imposing observational
constraints on the current values of $\alpha_{0}=\alpha(\mathbf{y}_{0})$, and
the matter parameter $\Omega_{0}=1-\Omega_{\phi}(\mathbf{y}_{0})$, restrict
the location of the present state of the universe, $\mathbf{y}_{0}$, in state
space. Finally in Section \ref{concl} we draw our conclusions and discuss
further possible extensions.

\section{Field equations}

\label{field}

In the teleparallel equivalence of general relativity one introduces a
non-holonomic frame by means of the functions $h_{i}^{\mu}$ so that the
tangent vectors to the new coordinates are the vectors $e_{i}=h_{i}^{\mu
}\left(  x\right)  \partial_{i}$ whose Lie bracket is $[e_{i},e_{j}%
]=c_{ij}^{...k}e_{k}$ where $c_{(ij)}^{...k}=0.$ In the nonholonomic
coordinates the connection is not symmetric and it is given by the expression
\begin{equation}
\chi_{jk}^{i}=\{_{jk}^{i}\}+\frac{1}{2}g^{ir}(c_{rj,k}+c_{rk,j}-c_{ij,r})
\label{lf.00}%
\end{equation}
where $c_{ijr}=g_{rk}c_{ij}^{...k}$ and $\{_{jk}^{i}\}$ is the standard
Riemannian connection. In case the vectors $e_{i}$ are orthonormal then they
form a vierbein field ${\mathbf{e}(x^{\mu})~}${the metric} becomes $\eta
_{ij},~$i.e. the Minkowski metric, while the connection coefficients reduce as
follows%
\begin{equation}
\chi_{jk}^{i}=\frac{1}{2}\eta^{ir}(c_{rj,k}+c_{rk,j}-c_{ij,r}). \label{lf.001}%
\end{equation}

In this case, the connection coefficients $\chi_{jk}^{i}$ are called the Ricci
rotation coefficients, also known as the Weitzenb\"{o}ck connection
\cite{sard01}. Defining $\chi_{ijk}=\eta_{ir}\chi_{jk}^{r}$ we find that
connection coefficients (\ref{lf.001}) have the property $\chi_{ijk}%
=-\chi_{jik};$ that is, they are antisymmetric in the first two indices.

The latter antisymmetric connection lead to the definition of the torsion
tensor%
\begin{equation}
T_{ijk}=\chi_{ijk}-\chi_{jik}, \label{lf.002}%
\end{equation}
and subsequently the quantities are defined%
\begin{equation}
K_{~~~\beta}^{\mu\nu}=-\frac{1}{2}({T^{\mu\nu}}_{\beta}-{T^{\nu\mu}}_{\beta
}-{T_{\beta}}^{\mu\nu}), \label{lf.04}%
\end{equation}
The quantities $K_{~~~\beta}^{\mu\nu}~$correspond to the contorsion tensor in
the case of torsion.

Assuming that the non-holonomic frame is inherent in the structure of
spacetime we have at our disposal $n^{2}$ new parameters $h_{i}^{\mu}$ which
can be used for the definition of dark energy. We associate with dark energy
the scalar field of geometric origin and noncanonical kinetic term with energy
momentum tensor \cite{an1,anbas}%
\begin{equation}
4\pi Ge\mathcal{T}_{a}^{\left(  \phi\right)  }{}^{\lambda}=\frac{1}{2}e\left(
h_{a}^{\sigma}\phi_{;\sigma}^{~~~;\lambda}-h_{a}^{\lambda}\phi^{;\mu\nu}%
g_{\mu\nu}\right)  -e\phi_{;\mu}S_{a}{}^{\mu\lambda}-\frac{1}{4}eV\left(
\phi\right)  h_{a}^{\lambda}. \label{lf.01}%
\end{equation}
where the geometric object ${S_{\beta}}^{\mu\nu}$ is defined as
\[
{S_{\beta}}^{\mu\nu}=\frac{1}{2}({K^{\mu\nu}}_{\beta}+\delta_{\beta}^{\mu
}{T^{\theta\nu}}_{\theta}-\delta_{\beta}^{\nu}{T^{\theta\mu}}_{\theta}).
\]

If $\mathcal{T}_{a}^{\left(  m\right)  }{}^{\lambda}$ is the energy momentum
tensor which describes the usual matter source then the gravitational field
equations~of teleparallel gravity are of second-order and are given by the
expression%
\begin{equation}
eG_{a}^{\lambda}=4\pi Ge\left(  \mathcal{T}_{a}^{\left(  m\right)  }%
{}^{\lambda}+\mathcal{T}_{a}^{\left(  \phi\right)  }{}^{\lambda}\right)  ,
\label{lf.02}%
\end{equation}
where$~G_{a}^{\lambda}$ is the Einstein-Tensor, where can be written with the
use of teleparallel quantities as
\begin{equation}
eG_{i}^{\rho}\mathbf{=}2\left(  e^{-1}\partial_{\mu}(ee_{i}^{\phantom{i}\rho
}S_{\rho}{}^{\mu\nu})+e_{i}^{\phantom{i}\lambda}T^{\rho}{}_{\mu\lambda}%
S_{\rho}{}^{\nu\mu}+\frac{1}{4}e_{i}^{\phantom{i}\rho}T\right)
\end{equation}
in which the scalar $T$ is defined as $T={S_{\beta}}^{\mu\nu}{T^{\beta}}%
_{\mu\nu}$.

Furthermore, we assume that the additional matter source is minimally coupled
with the scalar field the conservation equation (Bianchi identity) gives%
\begin{equation}
\left(  \mathcal{T}^{\left(  m\right)  a}{}^{\lambda}+\mathcal{T}^{\left(
\phi\right)  a}{}^{\lambda}\right)  _{;\lambda}=0\rightarrow\left(
\mathcal{T}^{\left(  m\right)  a}{}^{\lambda}\right)  _{;\lambda}=0\text{ ~and
~}\left(  \mathcal{T}^{\left(  \phi\right)  a}{}^{\lambda}\right)  _{;\lambda
}=0. \label{lf.05}%
\end{equation}

From (\ref{lf.02}) we observe that the field equations are written in the
Einstein frame. Furthermore, as far as the origin of $\phi$ is concerned, that
can have geometric origin and describe the higher-order terms of an extended
$f-$gravitational theory, for more details see \cite{an1,anbas} and
\cite{ftb1,ftb2}.

A well known analogue is the Brans-Dicke scalar field \cite{mod0}. Indeed the
latter when the Brans-Dicke parameter is zero, which corresponds to the
so-called O'Hanlon theory \cite{mod00}, is equivalent with the $f\left(
R\right)  $-gravity in the metric formalism \cite{Sotiriou}. In particular the
Brans-Dicke field attributes the higher-order derivatives of $f\left(
R\right)  $ gravity and the fourth-order theory can be written as second-order
theory by increasing at the same time the number of degrees of freedom. Hence,
in that explicitly analogue the energy momentum tensor (\ref{lf.01})
attributes the higher-order terms of a fourth-order $f-$theory in which the
invariant which is used for the modification of the Einstein-Hilbert action is
the boundary term which relates the Ricci scalar and the invariant $T$ of the
two connections in the holonomic and unholonomic frame, more details can be
found in \cite{an1,anbas}.

\subsection{Fourth-order theory of gravity}

Let us now discuss the variational problem which describes the gravitational
field equations (\ref{lf.02}), where $\mathcal{T}_{a}^{\left(  \phi\right)
}{}^{\lambda}$ is defined by (\ref{lf.01}).

Consider now the gravitational Action Integral to be%

\begin{equation}
S\equiv\frac{1}{16\pi G}\int d^{4}xe\left[  f(T,R+T)\right]  +S_{m}\equiv
\frac{1}{16\pi G}\int d^{4}xe\left[  f(T,B)\right]  +S_{m}, \label{ftb.01}%
\end{equation}
in which $e=\det(e_{\mu}^{i})=\sqrt{-g},~S_{m}$ is the Action Integral for the
matter source and $B$ is the boundary term $B=2e_{\nu}^{-1}\partial_{\nu
}\left(  eT_{\rho}^{~\rho\nu}\right)  $ which defined as
\begin{equation}
B=T+R
\end{equation}
where $R$ is the Ricciscalar. \ Gravitational actions of the form of
(\ref{ftb.01}) have bee considered previously in \cite{ftb1} and \cite{ftb2}.
As has been shown in \cite{ftb2} Action (\ref{ftb.01}) generalize $f\left(
T\right)  $-gravity while $f\left(  R\right)  $ gravity can be recovered.
Moreover, because of the second-derivative terms which are included in the
boundary $B$, the resulting gravitational field equations of (\ref{ftb.01})
are of fourth-order \cite{an1}.

Indeed, variation with respect to the vierbein field \ provides the field
equations \cite{ftb2}
\begin{align}
4\pi Ge\mathcal{T}_{a}^{\left(  m\right)  }{}^{\lambda}  &  =\frac{1}{2}%
ee_{a}^{\lambda}\left(  f_{,B}\right)  ^{;\mu\nu}g_{\mu\nu}-\frac{1}{2}%
ee_{a}^{\sigma}\left(  f_{,B}\right)  _{;\sigma}^{~~~;\lambda}+\frac{1}%
{4}e\left(  Bf_{,B}-\frac{1}{4}f\right)  e_{a}^{\lambda}\,+(eS_{a}{}%
^{\mu\lambda})_{,\mu}f_{,T}\nonumber\\
&  ~\ ~+e\left(  (f_{,B})_{,\mu}+(f_{,T})_{,\mu}\right)  S_{a}{}^{\mu\lambda
}~-ef_{,T}T^{\sigma}{}_{\mu a}S_{\sigma}{}^{\lambda\mu}, \label{ftb.02}%
\end{align}
where $\mathcal{T}_{a}^{\left(  m\right)  }{}^{\lambda}$ is the
energy-momentum tensor of the matter source.

We follow the analysis described in \cite{anbas} and we rewrite the field
equations (\ref{ftb.02}) as follows%
\begin{equation}
eG_{a}^{\lambda}=G_{eff}e\left(  \mathcal{T}_{a}^{\left(  m\right)  }%
{}^{\lambda}+\mathcal{T}_{a}^{\left(  DE\right)  }{}^{\lambda}\right)  ,
\label{ftb.07}%
\end{equation}
in which $G_{eff}=\frac{4\pi G}{f_{,T}}$ denotes the effective varying
\textquotedblleft gravitational constant\textquotedblright, and the
energy-momentum tensor $\mathcal{T}_{a}^{\left(  DE\right)  }{}^{\lambda}$ is
defined as \cite{anbas}
\begin{align}
4\pi Ge\mathcal{T}_{a}^{\left(  DE\right)  }{}^{\lambda}  &  =-\left[
\frac{1}{4}\left(  Tf_{,T}-f\right)  eh_{a}^{\lambda}+e(f_{,T})_{,\mu}S_{a}%
{}^{\mu\lambda}\right]  +\label{ftb.06}\\
&  -\left[  e(f_{,B})_{,\mu}S_{a}{}^{\mu\lambda}-\frac{1}{2}e\left(
e_{a}^{\sigma}\left(  f_{,B}\right)  _{;\sigma}^{~~~;\lambda}-e_{a}^{\lambda
}\left(  f_{,B}\right)  ^{;\mu\nu}g_{\mu\nu}\right)  +\frac{1}{4}%
eBe_{a}^{\lambda}f_{,B}\right]  ,\nonumber
\end{align}
which includes the fourth-order derivatives of the theory.

The latter energy-momentum tensor can be seen as the geometric dark-energy
source which drives the dynamics of the universe in order to explain the
acceleration phases of the universe, for discussions on geometric dark-energy
models see \cite{dark1,dark2} and references therein.

As we saw, in general the \textquotedblleft gravitational
constant\textquotedblright\ is varying with a function of $f^{-1}\left(
T\right)  $. However, if we assume now that $f\left(  T,B\right)  $ is a
linear function on $T$, that is $f_{,T}=const$, that is $f\left(  T,B\right)
=T+F\left(  B\right)  $, then we derive that $G_{eff}=4\pi G$. That simplest
scenario was studied for the first time in \cite{an1}. \

Moreover, in $T+F\left(  B\right)  $ theory, the geometric dark-energy
momentum tensor (\ref{ftb.06}) is simplified as \cite{anbas}
\begin{equation}
4\pi Ge\mathcal{T}_{a}^{\left(  B\right)  }{}^{\lambda}=-\left[
e(F_{,BB})B_{;\mu}S_{a}{}^{\mu\lambda}-\frac{1}{2}e\left(  e_{a}^{\sigma
}\left(  F_{,B}\right)  _{;\sigma}^{~~~;\lambda}-e_{a}^{\lambda}\left(
F_{,B}\right)  ^{;\mu\nu}g_{\mu\nu}\right)  +\frac{1}{4}e\left(
BF_{,B}-F\right)  e_{a}^{\lambda}\right]  \label{ftb.010}%
\end{equation}
or equivalently is written in the form of (\ref{lf.01}) where now the field
$\phi$ describes second-order terms, that is, $\phi=F\left(  B\right)  _{,B}%
$,~with $V\left(  \phi\right)  =$ $\left(  F-BF_{,B}\right)  .$~Therefore, the
geometric origin for the field $\phi$ is obvious. However, because
$G_{eff}=4\pi G=const$. we can say that the field $\phi$ is defined in the
Einstein frame, in contrary to the Scalar-tensor theories defined in the
Jordan frame.

That specific form of $f\left(  T,B\right)  $-theory it is possible to
provides cosmological eras which describes the two acceleration phases of our
universe, the inflation and the late-time acceleration \cite{an1}, while an
epoch where the geometric dark-energy fluid mimics an ideal gas can be
recovered \cite{anbas}. \ \ Another special property of the $T+F\left(
B\right)  $ theory is that when $B$ is constant, then the gravitational field
equations (\ref{ftb.02}) are those of General Relativity with cosmological constant.

Furthermore, as we shall below for the $T+F\left(  B\right)  ~$theory in a
FLRW and in a Bianchi I background it is possible to describe the field
equations by using the minisuperspace approach. Such a description is
important in order to apply mathematical methods from analytical mechanics and
derive analytical solutions for the field equations.

\subsection{FLRW}

In the case of an isotropic and homogeneous spacetime with zero spatial
curvature the line element is
\begin{equation}
ds^{2}=-N^{2}\left(  t\right)  dt^{2}+a^{2}\left(  t\right)  \left(
dx^{2}+dy^{2}+dz^{2}\right)  , \label{lf.07}%
\end{equation}
where $a\left(  t\right)  $ is the scale-factor of the three dimensional
Euclidean space and $N\left(  t\right)  $ is the lapse function. The commoving
observers are $u^{\mu}=\frac{1}{N}\delta_{t}^{\mu}.$~For the vierbein we
considered the quantities
\begin{equation}
h_{\mu}^{i}(t)=\mathrm{diag}\left(  N\left(  t\right)  ,a\left(  t\right)
,a\left(  t\right)  ,a\left(  t\right)  \right)  . \label{lf.06}%
\end{equation}

We assume that the matter source for the comoving observer is that of a
perfect fluid with matter density $\rho_{m}$ and pressure $p_{m};$ the energy
momentum tensor for the comoving observers is given by the expression
\begin{equation}
\mathcal{T}_{a\lambda}^{\left(  m\right)  }{}=\left(  \rho_{m}+p_{m}\right)
u_{a}u_{\lambda}+p_{m}g_{\alpha\lambda}, \label{lf.08}%
\end{equation}
while the Bianchi identity provides%
\begin{equation}
\dot{\rho}_{m}+3\frac{\dot{a}}{a}\left(  \rho_{m}+p_{m}\right)  =0.
\label{lf.09}%
\end{equation}
Furthermore, we consider that the equation of state parameter of the matter
source is constant, such that $p_{m}=w_{m}\rho_{m}$, and we impose the
restriction $w_{m}\in[-1,1]$; hence from (\ref{lf.09}) it follows $\rho
_{m}\left(  t\right)  =\rho_{m0}a\left(  t\right)  ^{-3\left(  w_{m}+1\right)
}.$

For the frame (\ref{lf.06}) from expression (\ref{lf.01}) we calculate that
the nonzero components of the energy momentum tensor $\mathcal{T}_{a}^{\left(
\phi\right)  }{}^{\lambda}$ are
\begin{equation}
4\pi Ge\mathcal{T}_{t}^{\left(  \phi\right)  }{}^{t}=-\left(  3\frac{\dot
{a}\dot{\phi}}{aN^{2}}+\frac{1}{2}V\left(  \phi\right)  \right)  \label{lf.11}%
\end{equation}
and%
\begin{equation}
4\pi Ge\mathcal{T}_{x}^{\left(  \phi\right)  }{}^{x}=4\pi Ge\mathcal{T}%
_{y}^{\left(  \phi\right)  }{}^{y}=4\pi Ge\mathcal{T}_{z}^{\left(
\phi\right)  }{}^{z}=-\left(  \frac{\ddot{\phi}}{N^{2}}-\frac{\dot{\phi}%
\dot{N}}{N^{3}}+\frac{1}{2}V\left(  \phi\right)  \right)  . \label{lf.12}%
\end{equation}

The nonzero components of the Einstein tensor are calculated to be
\begin{equation}
G_{0}^{0}=-3\left(  \frac{\dot{a}}{aN}\right)  ^{2}~,~G_{x}^{x}=G_{y}%
^{y}=G_{z}^{z}=-\left(  2\frac{\ddot{a}}{aN^{2}}+\left(  \frac{\dot{a}}%
{aN}\right)  ^{2}-2\frac{\dot{a}\dot{N}}{aN^{3}}\right)  . \label{lf.12b}%
\end{equation}

From expressions (\ref{lf.11}) and (\ref{lf.12}) for the comoving observer
$u^{\mu}$ we compute the energy density and the pressure of the field $\phi$
\ as follows%
\[
\rho_{\phi}=3\frac{\dot{a}\dot{\phi}}{aN^{2}}+\frac{1}{2}V\left(  \phi\right)
~,~p_{\phi}=-\left(  \frac{\ddot{\phi}}{N^{2}}-\frac{\dot{\phi}\dot{N}}{N^{3}%
}+\frac{1}{2}V\left(  \phi\right)  \right)
\]
and the equation of state parameter is%
\begin{equation}
w_{DE}=\frac{p_{\phi}}{\rho_{\phi}}=-\frac{N\ddot{\phi}-\dot{\phi}\dot
{N}+\frac{1}{2}N^{3}V\left(  \phi\right)  }{3N\frac{\dot{a}}{a}\dot{\phi
}+\frac{1}{2}N^{3}V\left(  \phi\right)  }. \label{lf.14}%
\end{equation}

Finally, the conservation equation (\ref{lf.05}) for the field $\phi$ gives%
\begin{equation}
\frac{1}{6}V_{,\phi}+\frac{\ddot{a}}{\dot{a}N^{2}}+2\left(  \frac{\dot{a}}%
{aN}\right)  ^{2}-\frac{\dot{a}\dot{N}}{aN^{3}}=0. \label{lf.15}%
\end{equation}

In the case where the lapse function $N\left(  t\right)  $ is constant, i.e.
$N\left(  t\right)  =1$, the gravitational field equations (\ref{lf.02}) take
the following simple form%
\begin{equation}
3H^{2}=3H\dot{\phi}+\frac{1}{2}V\left(  \phi\right)  +\rho_{m}, \label{lf.16}%
\end{equation}%
\begin{equation}
2\dot{H}+3H^{2}=\ddot{\phi}+\frac{1}{2}V\left(  \phi\right)  -p_{m},
\label{lf.17}%
\end{equation}
and the constraint equations are
\begin{equation}
\frac{1}{6}V_{,\phi}+\dot{H}+3H^{2}=0, \label{lf.18}%
\end{equation}%
\begin{equation}
\dot{\rho}_{m}+3H\left(  \rho_{m}+p_{m}\right)  =0, \label{lf.19}%
\end{equation}
where $H=\frac{\dot{a}}{a}$ is the Hubble function. Recall, that for arbitrary
lapse function the Hubble function is defined as $H\left(  t\right)
=\frac{\dot{a}}{aN}$.

\subsection{Minisuperspace description}

Following \cite{an1} we construct a point-like Lagrange so that the field
equations are derived from the Hamiltonian variational principle of least
action. The corresponding Lagrange function is
\begin{equation}
\mathcal{L}\left(  N,a,\dot{a},\phi,\dot{\phi}\right)  =-\frac{6}{N}a\dot
{a}^{2}+\frac{6}{N}a^{2}\dot{a}\dot{\phi}-Na^{3}V\left(  \phi\right)
-2\rho_{m0}Na^{-3w_{m}}, \label{lf.20}%
\end{equation}
which is a singular Lagrangian in the sense that the Hessian matrix
$\frac{\partial^{2}\mathcal{L}}{\partial\dot{a}\partial\dot{\phi}}$ vanishes.
This is to be expected because the field equations admit second-order
derivatives of the variables $\left(  a,\phi\right)  $, while the variable $N$
provides the constraint equation $G_{0}^{0}=T_{0}^{0}$.

Without loss of generality we may consider that $N=N\left(  a,\phi\right)  .$
Then the field equations describe the evolution of a canonical particle moving
in a two dimensional space under the action of an effective potential. From
the kinetic term of (\ref{lf.20}) we construct the minisuperspace metric
$\chi_{ij}=\frac{\partial^{2}\mathcal{L}}{\partial\dot{a}\partial\dot{\phi}%
}=\frac{\partial^{2}L}{\partial\dot{q}^{i}\partial\dot{q}^{j}}$, while the
effective potential is
\begin{equation}
V_{eff}\left(  a,\phi\right)  =Na^{3}V\left(  \phi\right)  +2\rho
_{m0}Na^{-3w_{m}}. \label{lf.21}%
\end{equation}

Finally the constraint equation (\ref{lf.16}) is the Hamiltonian invariant,
because the field equations are autonomous, which has a specific value.
Specifically, because of the constraint the Hamiltonian vanishes.

In the case in which $w_{m}=1$ the matter source is called stiff fluid of the
spacetime can it can be attributed to an extra degree of freedom, that is, it
corresponds to additional free scalar fields. That property is used to extend
our analysis in the case of the vacuum Bianchi I model. \

In the following, we derive the field equations for the case of the vacuum
Bianchi I universe and we show explicity how Lagrangian (\ref{lf.20})
describes the field equations for the Bianchi I universe when $\rho_{m0}$ is
related with the integration constants for the anisotropic parameters of the
inhomogeneous spacetime.

\subsection{Bianchi I}

In Bianchi I spacetime the line element in the Misner variables is written as
follows,%
\begin{equation}
ds^{2}=-N^{2}\left(  t\right)  dt^{2}+a^{2}\left(  t\right)  \left(
e^{-2\beta_{+}\left(  t\right)  }dx^{2}+e^{\beta_{+}\left(  t\right)
+\sqrt{3}\beta_{-}\left(  t\right)  }dy^{2}+e^{\beta_{+}\left(  t\right)
-\sqrt{3}\beta_{-}\left(  t\right)  }dy^{2}\right)  . \label{lf.22}%
\end{equation}
The latter line element admits a three dimensional abelian Killing group.
Functions $\beta_{+},~\beta_{-}$ are called the anisotropic parameters
\cite{rayan}.

We consider the diagonal frame%
\begin{equation}
h_{\mu}^{i}(t)=\mathrm{diag}(N\left(  t\right)  ,a(t)e^{-\beta_{+}\left(
t\right)  },a(t)e^{\frac{1}{2}\left(  \beta_{+}\left(  t\right)  +\sqrt
{3}\beta_{-}\left(  t\right)  \right)  },a(t)e^{\frac{1}{2}\left(  \beta
_{+}\left(  t\right)  -\sqrt{3}\beta_{-}\left(  t\right)  \right)  }),
\label{lf.23}%
\end{equation}
from where we calculate the invariant%
\begin{equation}
T=\left(  -6\left(  \frac{\dot{a}}{aN}\right)  ^{2}+\frac{3}{2}\left(
\frac{\dot{\beta}_{+}}{N}\right)  ^{2}+\frac{3}{2}\left(  \frac{\dot{\beta
}_{-}}{N}\right)  ^{2}\right)  \label{lf.24}%
\end{equation}
and the corresponding Lagrangian of the field equations (\ref{lf.02}) in the
case of the vacuum is derived to be%
\begin{equation}
\mathcal{L}\left(  N,a,\dot{a},\phi,\dot{\phi}\right)  =-\frac{6}{N}a\dot
{a}^{2}+\frac{3}{2}Na^{-3}\left(  \frac{a^{3}\dot{\beta}_{+}}{N}\right)
^{2}+\frac{3}{2}Na^{-3}\left(  \frac{a^{3}\dot{\beta}_{-}}{N}\right)
^{2}+\frac{6}{N}a^{2}\dot{a}\dot{\phi}-Na^{3}V\left(  \phi\right)  .
\label{lf.25}%
\end{equation}

For the Lagrangian (\ref{lf.25}) we observe that the quantities
\begin{equation}
\Phi_{+}=\left(  \frac{a^{3}\dot{\beta}_{+}}{N}\right)  \text{ and }\Phi
_{-}=\left(  \frac{a^{3}\dot{\beta}_{-}}{N}\right)
\end{equation}
are conservation laws, that is $\frac{d\Phi_{\pm}}{dt}=0$, which means that
with the application of the conservation laws the dynamical system can be
reduced to that of FLRW Lagrangian (\ref{lf.20}) where $\rho_{m0}=\rho
_{m0}\left(  \Phi_{+},\Phi_{-}\right)  $ and $w_{m}=1$.

\section{Cartan formalism and symmetries}

\label{cartan}

In this section we briefly discuss the method of Cartan for the study of
symmetries of Lagrange equations and consequently of the admitted conservation
laws. Because we are interested on systems of differential equations of
second-order we consider Lagrangians of the form $\mathcal{L}=\mathcal{L}%
\left(  t,x^{j},\dot{x}^{j}\right)  $ where $t$ is the independent variable,
$x^{i}\left(  t\right)  $ are the dependent variables and a dot denotes total
derivative with respect to $t$.

From the variation of the action $S=\int Ldt$, follows the Euler Lagrange
equations~$E_{L}\left(  L\right)  =0$ where $E_{\mathcal{L}}=\frac{d}{dt}%
\frac{\partial}{\partial\dot{x}^{i}}-\frac{\partial}{\partial x^{i}}~$is the
Euler-operator. Assume that the field equations are written in the form
\[
\ddot{x}^{i}=\Lambda^{i}\left(  t,x^{j},u^{j}\right)
\]
where $u^{i}=\dot{x}^{j}.$ We define the associated vector field $A$ \ to the
Lagrangian system, called the Hamiltonian flow, as follows
\[
A=\partial_{t}+u^{i}\partial_{i}+\Lambda^{i}\partial_{u^{i}}%
\]
where $\Lambda^{i}=\Lambda^{i}\left(  t,x^{j},u^{j}\right)  $ is defined by
the condition%
\begin{equation}
\frac{\partial^{2}\mathcal{L}}{\partial u^{i}\partial u^{j}}\Lambda^{j}%
=\frac{\partial\mathcal{L}}{\partial x^{i}}-\frac{\partial^{2}\mathcal{L}%
}{\partial u^{i}\partial x^{j}}u^{j}-\frac{\partial^{2}\mathcal{L}}{\partial
u^{i}\partial t}. \label{lf.27}%
\end{equation}

In the cotangent space we consider the basis
\begin{equation}
\left(  de^{1},de^{2},de^{3}\right)  =\left(  dx^{i}-u^{i}dt~,~du^{i}%
-\Lambda^{i}dt~,~dt\right)  , \label{lf.26b}%
\end{equation}
We note that the vector field $A=\partial_{t}+u^{i}\partial_{i}+\Lambda
^{i}\partial_{u^{i}}$ has the property\footnote{The operator $i_{A}$ denotes
the left-hook or antiderivation with respect the Hamiltonian flow.}
$i_{A}\left(  de^{1},de^{2},de^{3}\right)  =\left(  0,0,0\right)  .$ It is
easy to show that every closed differential form $df$ in that basis is
expressed as follows
\begin{equation}
df=\frac{\partial f}{\partial x^{i}}\left(  dx^{i}-u^{i}dt\right)
+\frac{\partial f}{\partial u^{i}}\left(  du^{i}-\Lambda^{i}dt\right)
+A\left(  f\right)  dt. \label{lf.27b}%
\end{equation}

We introduce the Cartan 1-form $\theta$ \cite{Sarlet}%
\begin{equation}
\theta=\mathcal{L}dt+\frac{\partial\mathcal{L}}{\partial u^{i}}\left(
dx^{i}-u^{i}dt\right)  . \label{lf.28}%
\end{equation}

$\theta$ is the the pullback under the Legendre transform of the fundamental
one form $u_{i}dx^{i}-Hdt$ in Hamiltonian mechanics where $H\ $is the
Hamiltonian. In terms of $\theta$ the equations of motions are
\begin{equation}
i_{A}\left(  d\theta\right)  =0. \label{lf.29a}%
\end{equation}

$d\theta$ is a 2-form called the second Cartan form which in basis
(\ref{lf.26b}) is expressed as follows%
\begin{equation}
d\theta=\frac{\partial\mathcal{L}}{\partial u^{i}\partial u^{j}}\left[
\left(  dx^{j}-u^{j}dt\right)  \curlywedge\left(  dx^{i}-u^{i}dt\right)
+\left(  du^{i}-\Lambda^{i}dt\right)  \curlywedge\left(  dx^{j}-u^{j}%
dt\right)  \right]  , \label{lf.30}%
\end{equation}
where $\curlywedge$ denotes the wedge product.

Therefore, if there exists a closed-form $f$, such that two Cartan one forms
are related such that%
\begin{equation}
\bar{\theta}-\theta=df, \label{lf.31}%
\end{equation}
then $\theta$ and $\bar{\theta}$ describe the same field equations, because by
definition $d\left(  \bar{\theta}-\theta\right)  =d^{2}f\equiv0$.

In the tangent space consider the point transformation
\begin{equation}
\left(  \bar{t},\bar{x}^{i},\bar{u}^{i}\right)  =\left(  t+\varepsilon
\xi\left(  t,x^{j},u^{j}\right)  ,x^{i}+\varepsilon\eta^{i}\left(
t,x^{j},u^{j}\right)  ,u^{i}+\varepsilon\zeta^{i}\left(  t,x^{j},u^{j}\right)
\right)  \label{lf.32}%
\end{equation}
generated by the vector field
\begin{equation}
X=\xi\frac{\partial}{\partial t}+\eta^{i}\frac{\partial}{\partial x^{i}}%
+\zeta^{i}\frac{\partial}{\partial u^{i}}. \label{lf.32a}%
\end{equation}

\bigskip We say that $X$ is a Cartan symmetry of the Lagrangian $L$ \ if
\begin{equation}
L_{X}\left(  d\theta\right)  =0
\end{equation}
where $L_{X}$ denotes the Lie derivative with respect to the vector field $X$.

Because $L_{X}\left(  d\theta\right)  =d\left(  L_{X}\theta\right)  $ we
conclude that the condition for a Cartan-symmetry is \cite{Crampin01}%
\begin{equation}
L_{X}\left(  d\theta\right)  =0\text{ or }L_{X}\left(  \theta\right)  =df.
\label{lf.33}%
\end{equation}
where $f$ is a function. The latter condition describes nothing else than
Noether's first theorem in the cotangent space while it is clear that $f$ is a
boundary term and inaccurately it is characterized as a gauged function.
Furthermore using the identity $L_{X}\theta=i_{X}\left(  d\theta\right)
+d\left(  i_{X}\theta\right)  $ we find that if $X$ is a Cartan symmetry then
$L_{X}\theta=0$ and
\begin{equation}
i_{A}d\left(  f-i_{X}\theta\right)  =0 \label{lf.33b}%
\end{equation}
which means that the quantity $\left(  f-i_{X}\theta\right)  $ is conserved.
Expression (\ref{lf.33b}) describes Noether's second theorem. Finally, if $X$
is a Cartan symmetry then $i_{\left[  A,X\right]  }d\theta=i_{A}L_{X}%
d\theta-L_{X}\left(  i_{A}d\theta\right)  $ which means that $L_{X}\left(
i_{A}d\theta\right)  =0$. The latter condition implies that Cartan symmetries
leave invariant the field equations (as expected) and form a subalgebra of the
Lie symmetries of the dynamical system.

\subsection{Symmetries for point-like Lagrangians}

For Lagrangian $\mathcal{L}=\frac{1}{2}g_{ij}\left(  x^{j}\right)  u^{i}%
u^{j}-V\left(  x^{k}\right)  $ the Cartan one-form is calculated to be
\begin{equation}
\theta=\left(  \frac{1}{2}g_{ij}u^{i}u^{j}-V\right)  dt+g_{ij}u^{j}\left(
dx^{i}-u^{i}dt\right)  \label{lf.35}%
\end{equation}
We consider the vector field $X=\xi\partial_{t}+\eta^{i}\partial_{i}+\zeta
^{i}\partial_{u^{i}}$ and require it to be a Cartan symmetry. However, since
$X$ is also a Lie symmetry it follows that $\zeta^{i}$ is not independent but
it can expressed explicitly in terms of $\xi,~\eta^{i}$ and their
derivatives\footnote{This is possible only for holonomic dynamical systems.}
\begin{equation}
\zeta^{i}=\eta_{,t}^{i}+u^{k}\eta_{,k}^{i}+\Lambda^{k}\eta_{,u^{k}}^{i}%
-u^{i}\left(  \xi_{,t}+u^{k}\xi_{,k}+\Lambda^{k}\xi_{,u^{k}}\right)  .
\label{lf.36}%
\end{equation}

From the symmetry condition (\ref{lf.33}) it follows
\begin{align*}
L_{X}\theta &  =\left[  g_{ij,k}\eta^{k}u^{j}+g_{ij}\zeta^{j}\right]
dx^{i}+g_{ij}u^{j}\eta_{,k}^{i}dx^{k}-\frac{1}{2}g_{ij}u^{i}u^{j}\xi
_{,k}dx^{k}+\\
&  +\left[  -g_{ij}u^{i}\zeta^{j}-\frac{1}{2}g_{ij,k}\eta^{k}u^{i}u^{j}%
+g_{ij}u^{j}\eta_{,t}^{i}-\frac{1}{2}g_{ij}u^{i}u^{j}\xi_{,t}-\eta^{k}%
V_{k}\right]  dt+\\
&  +\left[  g_{ij}u^{j}\eta_{,u^{k}}^{i}-\frac{1}{2}g_{ij}u^{i}u^{j}%
\xi_{,u^{k}}^{i}\right]  du^{k}\\
&  =f_{,t}dt+f_{,i}dx^{i}+f_{,u^{k}}du^{k}%
\end{align*}
from where we find the following set of equations
\begin{align}
\left[  g_{ij,k}\eta^{k}u^{j}+g_{ij}\zeta^{j}\right]  \delta_{r}^{i}%
+g_{ij}u^{j}\eta_{,k}^{i}\delta_{r}^{k}-\frac{1}{2}g_{ij}u^{i}u^{j}\xi
_{,k}\delta_{r}^{k}-f_{,r}  &  =0.\label{lf.37}\\
-g_{ij}u^{i}\zeta^{j}-\frac{1}{2}g_{ij,k}\eta^{k}u^{i}u^{j}+g_{ij}u^{j}%
\eta_{,t}^{i}-\frac{1}{2}g_{ij}u^{i}u^{j}\xi_{,t}-f_{,t}-\eta^{k}V_{k}  &
=0.\label{lf.38}\\
g_{ij}u^{j}\eta_{,u^{k}}^{i}-\frac{1}{2}g_{ij}u^{i}u^{j}\xi_{,u^{k}}%
^{i}-f_{,u^{k}}  &  =0. \label{lf.39}%
\end{align}

Specific forms of the functions $\xi$ and $\eta^{i}$ reduce the above system
to various special forms.

\subsubsection{Point transformations}

In this case both \ $\xi$ and~$\eta$ are independent of $u^{i}$ and the
resulting symmetry conditions become
\begin{align}
V_{,k}\eta^{k}+V\xi_{,t}+f_{,t}  &  =0~~,~~\eta_{,t}^{i}g_{ij}-\xi
_{,j}V-f_{,j}=0\label{lf.40}\\
L_{\eta}g_{ij}-\xi_{,t}g_{ij}  &  =0~~,~~\xi_{,i}=0~~,~~f_{,u^{k}}=0
\label{lf.41}%
\end{align}
whose general solution can be found in \cite{tsampandgrg}. \ Finally, the
corresponding conservation law is linear in the velocity $u^{i}$.

\subsubsection{Higher-order symmetries}

When \ $\xi$ and~$\eta$ are functions $u^{i}$ these symmetries are called
higher-order symmetries. A\ particular class of higher order symmetries are
the contact symmetries in which $\xi$ and $\eta^{i}$ are linear functions of
$u^{i}.$ In this case it can be shown that without loss of generality one can
set $\xi=0$ \ and $\eta^{i}=K_{~j}^{i}\left(  t,x^{k}\right)  u^{j}$. If this
is done then the symmetry conditions take the simple form
\begin{align}
K_{\left(  ij\right)  ,t}  &  =0~~,~~K_{(is;j)}=0\label{lf.42}\\
g_{ij}V^{k}K_{k}^{j}+f_{,i}  &  =0~~,~~f_{,t}=0~,~~f_{,u^{k}}=0. \label{lf.43}%
\end{align}

In a similar way the symmetry conditions can be derived for other dependence
of $\eta^{i}$ on $u^{i}$. In general it is easy to show that if $\eta^{i}$ is
a polynomial of rank $n$ on $u^{i}$ then the corresponding symmetry conditions
are polynomials of rank $\left(  n+1\right)  $ on $u^{i}$.%
\begin{align}
K_{\left(  irs;j\right)  }  &  =0~~,~~K_{\left(  irs\right)  ,t}%
=0~~\label{lf.44}\\
V^{k}K_{(iks)}  &  =0~~,~~f=0. \label{lf.45}%
\end{align}

While the point symmetries form a Lie algebra for the higher-order symmetries
there are some differences. For instance, if $K_{\left(  1\right)
},~K_{\left(  2\right)  }$ are second-rank tensor which produce two contact
symmetries with boundary terms $f_{\left(  1\right)  }$ and $f_{\left(
2\right)  }$ then it follows that%
\begin{equation}
\left[  K_{\left(  1\right)  },K_{\left(  2\right)  }\right]  _{SN}\left(
V_{k}\right)  +\left[  f_{\left(  1\right)  ;i},f_{\left(  2\right)
;j}\right]  =K^{ijk}V_{k}+\left[  f_{\left(  1\right)  ;i},f_{\left(
2\right)  ;j}\right]  \label{lf.46}%
\end{equation}
where $\left[  K_{\left(  1\right)  },K_{\left(  2\right)  }\right]  _{SN}$
denotes the Schouten-Nijenhuis Bracket. It follows that $K^{ijk}$ produces a
quadratic higher-order symmetry if and only if $\left[  f_{\left(  1\right)
;i},f_{\left(  2\right)  ;j}\right]  =0$. Therefore the commutator of two
higher order symmetries produces a higher-order symmetry of higher rank.

In the special case that the boundary terms for contact or higher symmetries
are zero, that is $f_{\left(  1\right)  }=f_{\left(  2\right)  }=0~$, or for
noncontact higher-order symmetries, it follows that if the Lagrangian
$\mathcal{L}$ ~admits the conserved quantities $I_{A}$,~$I_{B}$, defined as%
\begin{equation}
I_{A}=K_{ij_{1}...j_{a}}u^{i}u^{j_{1}}...u^{j_{a}}~,~I_{B}=T_{ij_{1}...j_{b}%
}u^{i}u^{j_{1}}...u^{j_{b}} \label{lf.47}%
\end{equation}
with $a,b>1,$ then also the quantity
\begin{equation}
I_{S}=S_{ij_{1}...j_{c}}u^{i}u^{j_{1}}...u^{j_{c}}~,~c=a+b-1 \label{lf.48}%
\end{equation}
is a conserved quantity, where $\mathbf{S\equiv}\left[  \mathbf{K,T}\right]
_{SN}$.

In the following Section, we continue with the determination of the unknown
parameters of the field equations (\ref{lf.02}) so that Cartan symmetries are
admitted which provide conserved quantities sufficient to prove the
integrability of the field equations and when it is feasible to write the
solution in closed-form.

\section{Symmetries and analytic Solutions}

\label{cartan2}

In the Lagrangian of the field equations (\ref{lf.20})\ without loss of
generality we consider that $N\left(  t\right)  =a^{3w_{m}}$. With that
selection the fluid term has been absorbed in the minisuperspace which
simplifies our calculations. Then the Cartan 1-form which describes the field
equations is calculated to be%
\begin{align}
\theta &  =\left(  -6a^{1-3w_{m}}\dot{a}^{2}+6a^{2-3w_{m}}\dot{a}\dot{\phi
}-a^{3\left(  1+w_{m}\right)  }V\left(  \phi\right)  +2\rho_{m0}\right)
dt+\nonumber\\
&  +\left(  -12a^{1-3w_{m}}\dot{a}+6a^{2-3w_{m}}\dot{\phi}\right)  \left(
da-\dot{a}dt\right)  +\left(  6a^{2-3w_{m}}\dot{a}\right)  \left(  d\phi
-\dot{\phi}dt\right)  , \label{lf.49}%
\end{align}
and the corresponding Hamiltonian flow is
\begin{align}
A  &  =\frac{\partial}{\partial t}+\dot{a}\frac{\partial}{\partial a}%
+\dot{\phi}\frac{\partial}{\partial\phi}+\left(  \left(  3w-2\right)
\frac{\dot{a}^{2}}{a}+\frac{a^{1+6w}}{6}V_{,\phi}\right)  \partial_{\dot{a}%
}+\nonumber\\
&  ~~~~~~~~~+\left(  3\left(  w-1\right)  \left(  \frac{\dot{a}}{a}\right)
^{2}+\frac{a^{6w}(3\left(  w+1\right)  V+2V_{,\phi})}{6}\right)
\partial_{\dot{\phi}}. \label{lf.50}%
\end{align}

The field equations have four degrees of freedom $\left(  a,\dot{a},\phi
,\dot{\phi}\right)  $, with the constraint equation, the conservation law of
\textquotedblleft energy\textquotedblright\ to be zero, that is the
Hamiltonian $H=0$. Hence, the determination of a second conservation law is
sufficient to prove the integrability of the field equations, as defined by
Liouville\footnote{In the following, with the term integrability we mean
Liouville integrability.}. Recall, that the second conservation law should be
linearly independent from the Hamiltonian and in involution.

Another important question is which kind of symmetries have to be used for the
determination of integrable systems. For instance there are different systems
which admit point symmetries and other higher-order symmetries. However these
two sets of systems are not independent and the systems which admit point
symmetries are included in the systems which admit higher-order symmetries.
This is easy to show by using the inverse problem to construct the symmetry
vector from the conservation law.

The inverse problem says that if $\Phi$ is a constant of motion for a system
with Cartan 1-form $\theta$, then there exists a vector field $X$ such that
$L_{X}\theta=d\left(  F+i_{X}\theta\right)  $, that is, $X$ is a Cartan
symmetry, while any vector field $Y=X+\lambda A$ is also a Cartan symmetry
which produces the same conservation law. Hence, all point symmetries generate
conservation laws linear in the velocities. However any function of a
conservation law is also a conservation law which means that if $\Phi_{P}$ is
a conservation law generated by a point symmetry then $\Phi_{C}=\left(
\Phi_{P}\right)  ^{2}$ is a conservation law associated with a contact
symmetry. \ Hence, we prefer to work with the higher-order symmetries and
specifically with the Cartan symmetries generated by contact transformations.
We omit the calculations for the derivation of Cartan symmetries and their
corresponding conservation laws and we continue with the direct presentation
of the results.

\subsection{Classification of Cartan symmetries}

We find that there are some differences between the potentials and the
conservation laws for $w_{m}\neq1$ and $w_{m}=1$.

\subsubsection{Non stiff fluid $w_{m}\neq1$}

\label{Sect:4.1.1}

In particular for $w_{m}\neq1$ we find that the scalar field potentials for
which the field equations admit Cartan symmetries generated by the contact
transformations are the following
\begin{equation}
V_{A}\left(  \phi\right)  =V_{1}\phi+V_{0}, \label{lf.51}%
\end{equation}%
\begin{equation}
V_{B}\left(  \phi\right)  =V_{1}e^{-3(w_{m}+1)\phi}+V_{2}e^{-6w_{m}\phi}.
\label{lf.52}%
\end{equation}
and%
\begin{equation}
V_{C}\left(  \phi\right)  =V_{1}e^{-3\left(  1+w_{m}\right)  \phi}%
+V_{2}e^{-\frac{3}{2}\left(  3+w_{m}\right)  } \label{lf.53}%
\end{equation}

The conservation law which corresponds to the potential $V_{A}\left(
\phi\right)  $ is
\begin{equation}
I_{A}=a^{4-6w_{m}}\dot{a}^{2}+\frac{V_{1}}{18}a^{6} \label{lf.54}%
\end{equation}
while for the potential $V_{B}\left(  \phi\right)  ~$the additional
conservation law is
\begin{equation}
I_{B}=\left(  \frac{\dot{a}-a\dot{\phi}}{a}\right)  ^{2}+\frac{(w_{m}%
-1)}{6w_{m}}V_{2}\left(  ae^{-\phi}\right)  ^{6w_{m}}~\text{for}~w_{m}\neq0,
\label{lf.55}%
\end{equation}
or%
\begin{equation}
I_{B}^{0}=\left(  \frac{\dot{a}}{a}-\dot{\phi}\right)  ^{2}-V_{2}\left(  \ln
a-\phi\right)  ~\text{\ for}~w_{m}=0. \label{lf.56}%
\end{equation}

For the potential $V_{C}\left(  \phi\right)  ~$the extra conservation law is%
\begin{align}
I_{C}  &  =a^{1-3w_{m}}\left(  2+3\left(  1-w_{m}\right)  \left(  \phi-\ln
a\right)  \right)  \dot{a}^{2}-3a^{2-3w_{m}}\left(  1+\left(  1-w_{m}\right)
\left(  \phi-\ln a\right)  \right)  \dot{a}\dot{\phi}+\nonumber\\
&  ~+a^{3\left(  1-w_{m}\right)  }\dot{\phi}^{2}+\frac{V_{1}}{6\left(
1+w_{m}\right)  }\left(  2+3\left(  w_{m}^{2}-1\right)  \left(  \ln
a-\phi\right)  \right)  a^{3\left(  1+w_{m}\right)  }e^{-3\left(
1+w_{m}\right)  \phi}+\nonumber\\
&  +\frac{V_{2}}{6}\left(  1+3\left(  1-w_{m}\right)  \left(  \phi-\ln
a\right)  \right)  a^{3\left(  1+w_{m}\right)  }e^{-\frac{3}{2}\left(
3+w_{m}\right)  \phi}. \label{lf.57}%
\end{align}

\subsubsection{Stiff fluid $w_{m}=1$}

\label{sect:4.1.2}

When $w_{m}=1$, that is, the matter source is that of stiff fluid the symmetry
analysis provides us with the potential
\begin{equation}
V_{D}=V_{1}e^{-3\phi}, \label{lf.58}%
\end{equation}
and $V_{A}\left(  \phi\right)  $ of expression (\ref{lf.51}).

The corresponding conservation laws are $I_{A}$ given by expression
(\ref{lf.54}) and
\begin{equation}
I_{D}=\left(  2\frac{\dot{a}}{a}-\dot{\phi}\right)  ^{2}-\frac{2}{3}V_{1}%
a^{6}e^{-3\phi}, \label{lf.60}%
\end{equation}

We proceed with the construction of the analytical solutions for the field equations.

\subsection{Analytic solutions}

There are various ways to describe the solution of \ a system of differential
equations. Usually when we refer to a solution we mean that there exists an
explicit formula which relates the dependent and the independent variables. If
that formula admits free parameters less from the number of degrees of freedom
of the system, the solution is characterized as a special solution, because it
is a solution for specific families of initial conditions.

However the existence of an explicit formula, that is a closed-form solution,
it is not always possible. For instance the solution of the well-known Abel
equation cannot be written always in \ closed-form expression. Another context
to express the solution of a dynamical system is to find the equivalent
reduced system. If the latter system can be integrated by quadratures then we
can construct the closed-form solution for the original system; however, in
general this is not possible.

Concerning our problem we have to reduce the field equations to a system of
two-first order equations. Indeed the constraint equation and the conservation
laws that we determined are sufficient to be described as the solution of the
field equations. However in order to simplify the expressions we follow the
method of Hamilton-Jacobi, for more details on the Hamilton-Jacobi method of
two-dimensional systems see \cite{daskaloyiannis,int2}.

\subsubsection{Potential $V_{A}\left(  \phi\right)  $}

For the potential $V_{A}\left(  \phi\right)  $ we prefer to work with the
equation of motions. In order to simplify the equations we perform the
coordinate transformation
\begin{equation}
a=e^{\chi}~,~\phi=\chi+\psi\label{lf.61}%
\end{equation}
and the field equations become%
\begin{equation}
6e^{3\left(  1-w_{m}\right)  \chi}\dot{\chi}\dot{\psi}-e^{3\left(
1+w_{m}\right)  \chi}\left(  V_{0}+V_{1}\left(  \chi+\psi\right)  \right)
-2\rho_{m0}=0 \label{lf.62}%
\end{equation}%
\begin{equation}
\ddot{\chi}+3\left(  1-w_{m}\right)  \dot{\chi}^{2}+V_{1}e^{6w_{m}\chi}=0
\label{lf.63}%
\end{equation}%
\begin{equation}
\ddot{\psi}+\frac{1}{6}e^{3w_{m}\chi}\left(  V_{1}+3\left(  1+w_{m}\right)
\left(  V_{0}+V_{1}\left(  \chi+\psi\right)  \right)  \right)  =0.
\label{lf.64}%
\end{equation}

We focus on equation (\ref{lf.63}) which is that of the scale factor. We see
that it can be written as follows
\begin{equation}
\dot{\chi}^{2}=\frac{1}{3}e^{-6\left(  1-w_{m}\right)  \chi}\left(  3\chi
_{0}-V_{1}e^{6\chi}\right)  , \label{lf.65}%
\end{equation}

Hence the explicit form of the Hubble function can be calculated, that is,
\begin{equation}
H^{2}\left(  a\right)  =a^{-6w_{m}}\left(  \frac{\dot{a}}{a}\right)
^{2}=\left(  \chi_{0}a^{-6}-\frac{V_{1}}{3}\right)  . \label{lf.66}%
\end{equation}
which means that $V_{1}$ is the cosmological constant and the integration
constant $\chi_{0}$ is the energy density of the stiff fluid which is
introduced by the theory. Recall that the Hubble function in general is
defined as $H\left(  t\right)  =\frac{1}{N}\frac{\dot{a}}{a}$

\subsubsection{Potential $V_{B}\left(  \phi\right)  $}

For the second potential, namely $V_{B}\left(  \phi\right)  $, we select the
new coordinates%
\begin{equation}
\phi=\ln a+\psi\label{lf.68}%
\end{equation}
where the solution of the Hamilton Jacobi equation for $w_{m}\neq0$ is given
as follows
\begin{align}
S\left(  a,\psi\right)   &  =\frac{\sqrt{6}a^{3\left(  1-w_{m}\right)  }%
\sqrt{V_{2}\left(  1-w_{m}\right)  e^{-6w_{m}\psi}+6w_{m}I_{B}}}{3\sqrt{w_{m}%
}\left(  1-w_{m}\right)  }\nonumber\\
&  -\sqrt{6w_{m}}\int\frac{2\rho_{m0}e^{3w_{m}\psi}+V_{1}e^{-3\psi}}%
{\sqrt{6w_{m}I_{B}e^{6w_{m}\psi}+V_{2}\left(  1-w_{m}\right)  }}d\psi
~,~w_{m}\neq0 \label{lf.69}%
\end{align}
which is clear that the system is supported by a Lie surface.

Furthermore in the new coordinates the reduced system is
\begin{equation}
a^{2-3w_{m}}\dot{a}=\frac{p_{\psi}}{6}~,~a^{2-3w_{m}}\dot{\psi}=\frac{p_{a}%
}{6a^{2}}. \label{lf.69a}%
\end{equation}

In the special case in which the matter source is dust, that is, $w_{m}=0$,
the solution of the Hamilton Jacobi equation takes the simplest form
\begin{equation}
S\left(  a,\psi\right)  =2a^{3}\sqrt{\left(  I_{B}^{0}-V_{2}\psi\right)
}+\frac{4\rho_{m0}}{V_{2}}\sqrt{\left(  I_{B}^{0}-V_{2}\psi\right)  }+\frac
{2}{\sqrt{3V_{2}}}D\left(  \sqrt{\frac{3\left(  I_{B}^{0}-V_{2}\psi\right)
}{V_{2}}}\right)  \label{lf.70}%
\end{equation}
where $D\left(  x\right)  $ is the Dawson function,~$D\left(  x\right)
=e^{-x^{2}}\int e^{x^{2}}dx^{2}$.

For $w_{m}=0$,\ that is $N\left(  t\right)  =1$, the reduced system
(\ref{lf.69a}) with the use of (\ref{lf.70}) becomes
\begin{equation}
6a^{2}\dot{a}=-\frac{2\rho_{m0}+V_{1}e^{-3\psi}}{\sqrt{\left(  I_{B}^{0}%
-V_{2}\psi\right)  }}~,~\dot{\psi}=\sqrt{\left(  I_{B}^{0}-V_{2}\psi\right)
}\label{lf.71}%
\end{equation}
that is
\begin{equation}
\psi\left(  t\right)  =\frac{I_{B}^{0}}{V_{2}}-\frac{V_{2}}{4}\left(
t-t_{0}\right)  ^{2}\label{lf.72}%
\end{equation}
and
\begin{equation}
a^{3}=\frac{2\rho_{m0}}{\left(  -V_{2}\right)  }\ln t+\frac{1}{2}\frac
{V_{1}e^{-3\frac{I_{B}^{0}}{V_{2}}}}{\left(  -V_{2}\right)  }\operatorname{Ei}%
\left(  \frac{3}{4}\left(  -V_{2}\right)  t^{2}\right)  \label{lf.73}%
\end{equation}
where $\operatorname{Ei}\left(  t\right)  $ is the exponential integral
function. Finally we see that when $\operatorname{Ei}\left(  t\right)  $
dominates the scale factor is approximated by the term%
\begin{equation}
a\left(  t\right)  \simeq\exp\left(  a_{1}t^{2}\right)  .\label{lf.74}%
\end{equation}

\subsubsection{Potential $V_{C}\left(  \phi\right)  $}

We perform the coordinate transformation%
\begin{equation}
a=r^{\frac{1}{3\left(  1-w_{m}\right)  }}~,~\phi=\frac{1}{3\left(
1-w_{m}\right)  }\ln\left(  r\right)  +\psi\label{lf.75}%
\end{equation}
which gives the Hamilton Jacobi equation%
\begin{equation}
\left(  w_{m}-1\right)  \left(  \frac{\partial S}{\partial r}\right)  \left(
\frac{\partial S}{\partial\psi}\right)  -4\rho_{m0}-2V_{1}e^{-3\left(
1+w_{m}\right)  \psi}-\frac{2V_{2}}{\sqrt{r}}e^{-\frac{3}{2}\left(
3+w_{m}\right)  \psi}=0. \label{lf.76}%
\end{equation}
where now it follows
\begin{equation}
\dot{r}=\frac{1}{2}\left(  1-w_{m}\right)  p_{\psi}~,~\dot{\psi}=\frac{1}%
{2}\left(  1-w_{m}\right)  p_{r}. \label{lf.77}%
\end{equation}

Hence, the action is calculated to be%
\begin{align}
S\left(  u,v\right)   &  =\frac{2\sqrt{2r}e^{-\frac{3}{2}\left(
1+w_{m}\right)  \psi}\sqrt{V_{1}\left(  1-w_{m}\right)  +2e^{3\left(
1+w\right)  \psi}\left(  1+w_{m}\right)  \left(  3I_{C}-\left(  1-3\psi\left(
w-1\right)  \right)  \rho_{m0}\right)  }}{\sqrt{3}\sqrt{\left(  1+w_{m}%
\right)  \left(  w_{m}-1\right)  }}+\nonumber\\
&  +\int\frac{6V_{2}\sqrt{\left(  1+w\right)  }e^{\frac{3}{2}\left(
1+w_{m}\right)  \psi}}{\sqrt{6}\sqrt{V_{1}\left(  1-w_{m}\right)
+2e^{3\left(  1+w\right)  \psi}\left(  1+w_{m}\right)  \left(  3I_{C}-\left(
1-3\psi\left(  w-1\right)  \right)  \rho_{m0}\right)  }}d\psi. \label{lf.78}%
\end{align}

In the special limit in which $\rho_{m0}=0$ and $I_{BC}=0$, the action takes
the simplest form%
\begin{equation}
S\left(  u,v\right)  =\frac{2\sqrt{2V_{1}r}e^{-\frac{3}{2}\psi\left(
1+w_{m}\right)  }}{3\left(  1-w_{m}^{2}\right)  }-\frac{\sqrt{2}\left(
1+w_{m}\right)  V_{2}e^{-3\psi}}{^{\sqrt{3V_{1}\left(  1-w_{m}^{2}\right)  }}%
}. \label{lf.79}%
\end{equation}
From the latter action and for dust fluid we take the following reduced system%
\begin{equation}
\dot{r}=\frac{\sqrt{6}V_{2}}{2\sqrt{V_{1}}}e^{-3\psi}-\sqrt{6}e^{-\frac{3}%
{2}\psi}\sqrt{V_{1}r}~,~\dot{\psi}=\frac{\sqrt{2V_{1}}}{\sqrt{r}}e^{-\frac
{3}{2}\psi} \label{lf.80}%
\end{equation}
from where we can see that for large values of $r$, that is for large value of
the scale factor $\dot{\psi}\simeq0$, which means that
\begin{equation}
\dot{r}\simeq c_{1}-c_{2}\sqrt{r}, \label{lf.81}%
\end{equation}
that is, the Hubble function is approximated by the closed-form expression
\begin{equation}
H\left(  a\right)  \simeq\left(  c_{1}a^{-3(1-w_{m})}-c_{2}a^{\frac
{-3(1-w_{m})}{2}}\right)  . \label{lf.82}%
\end{equation}

\subsubsection{Potential $V_{D}\left(  \phi\right)  $}

For the last potential and for $w_{m}=1$, we find that the normal coordinates
are
\begin{equation}
a=e^{\chi}~~,~~\phi=2\chi+\psi~ \label{lf.83}%
\end{equation}
where the reduced system takes the form
\begin{equation}
\dot{\chi}=\frac{p_{\psi}}{6}~,~\dot{\psi}=\frac{p_{\chi}-2p_{\psi}}{6}.
\label{lf.84}%
\end{equation}

The corresponding Hamilton-Jacobi equation is%
\begin{equation}
\left(  \frac{\partial S}{\partial\psi}\right)  ^{2}-\left(  \frac{\partial
S}{\partial\chi}\right)  \left(  \frac{\partial S}{\partial\psi}\right)
-6e^{-3\psi}V_{1}-12\rho_{m0}=0, \label{lf.85}%
\end{equation}
while the conservation law%
\begin{equation}
\left(  \frac{\partial S}{\partial\psi}\right)  ^{2}=36\bar{I}_{D}.
\label{lf.86}%
\end{equation}

From the above we find the action to be%
\begin{align}
S\left(  \chi,\psi\right)   &  =\sqrt{\bar{I}_{D0}}\chi+\frac{\sqrt{\bar
{I}_{D}}}{2}\psi+\frac{\sqrt{\left(  \bar{I}_{D}-48\rho_{m0}\right)
+24V_{1}e^{-3\psi}}}{3}+\nonumber\\
&  -\frac{\sqrt{\left(  \bar{I}_{D}-48\rho_{m0}\right)  }}{3}\arctan h\left(
\frac{\sqrt{\left(  \bar{I}_{D}-48\rho_{m0}\right)  +24V_{1}e^{-3\psi}}}%
{\sqrt{\left(  \bar{I}_{D}-48\rho_{m0}\right)  }}\right)  . \label{lf.87}%
\end{align}

The solution of the field equations can be written in closed form and the
scale factor is determined to be
\begin{equation}
a\left(  t\right)  =a_{0}e^{\omega_{0}t}\left(  1-6V_{1}\omega_{1}%
e^{-3\omega_{1}t}\right)  ^{-\frac{1}{3}}\label{lf.88}%
\end{equation}
where $\omega_{0,}~\omega_{1}$ are related with $\bar{I}_{D}$ and $\rho_{m0}$.
From the above solution, we observe that at late time the solution is
exponential, $a\left(  t\right)  \simeq a_{0}e^{\omega_{0}t}$ for $\omega
_{1}>0,$ and for $N\left(  t\right)  =$const., that is,  $w_{m}=0$, the future
solution it corresponds to the de Sitter universe while $\Omega_{m0}=0$.

\begin{figure}[ptb]
\includegraphics[height=6.5cm]{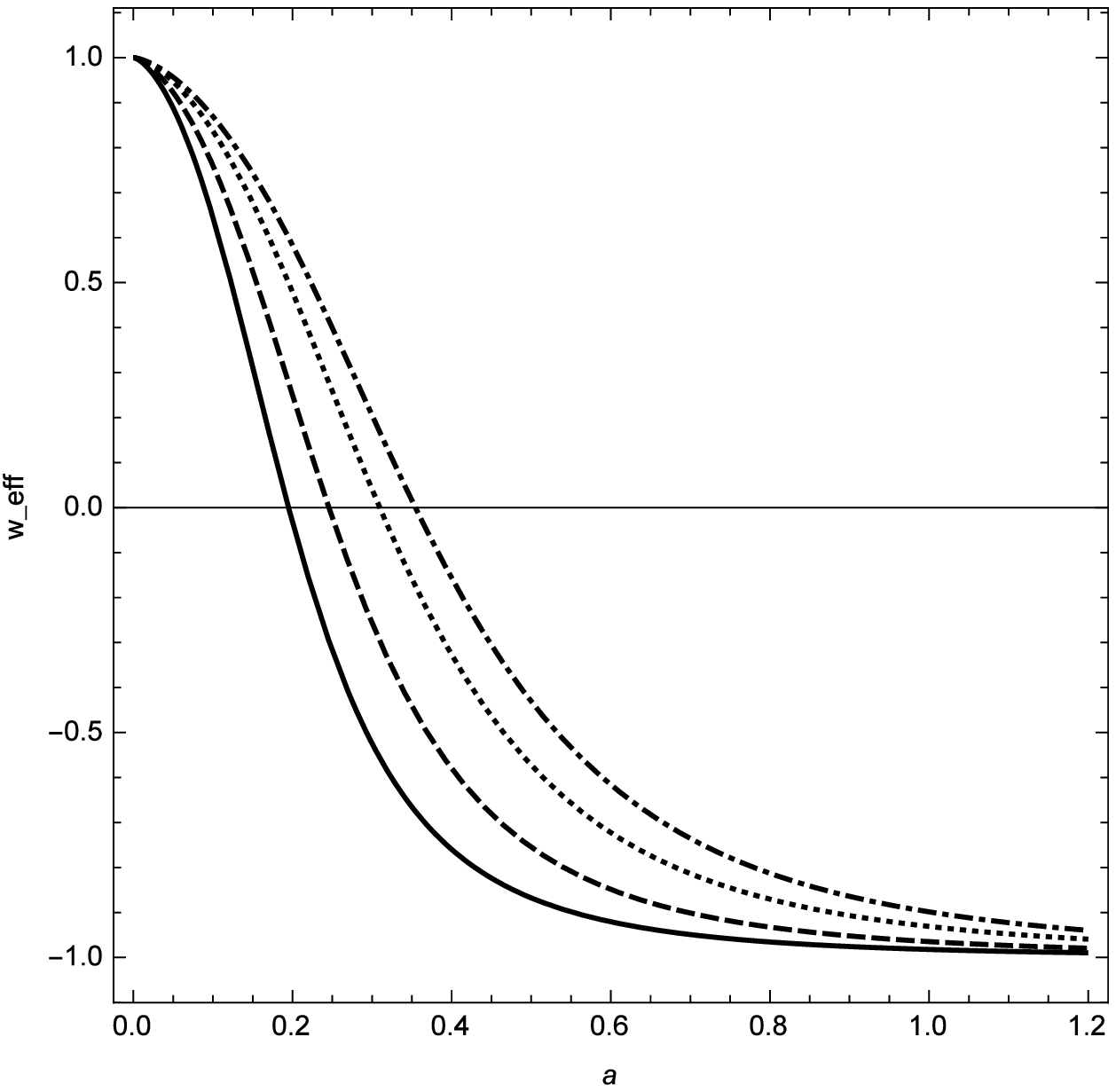}\centering
\includegraphics[height=6.5cm]{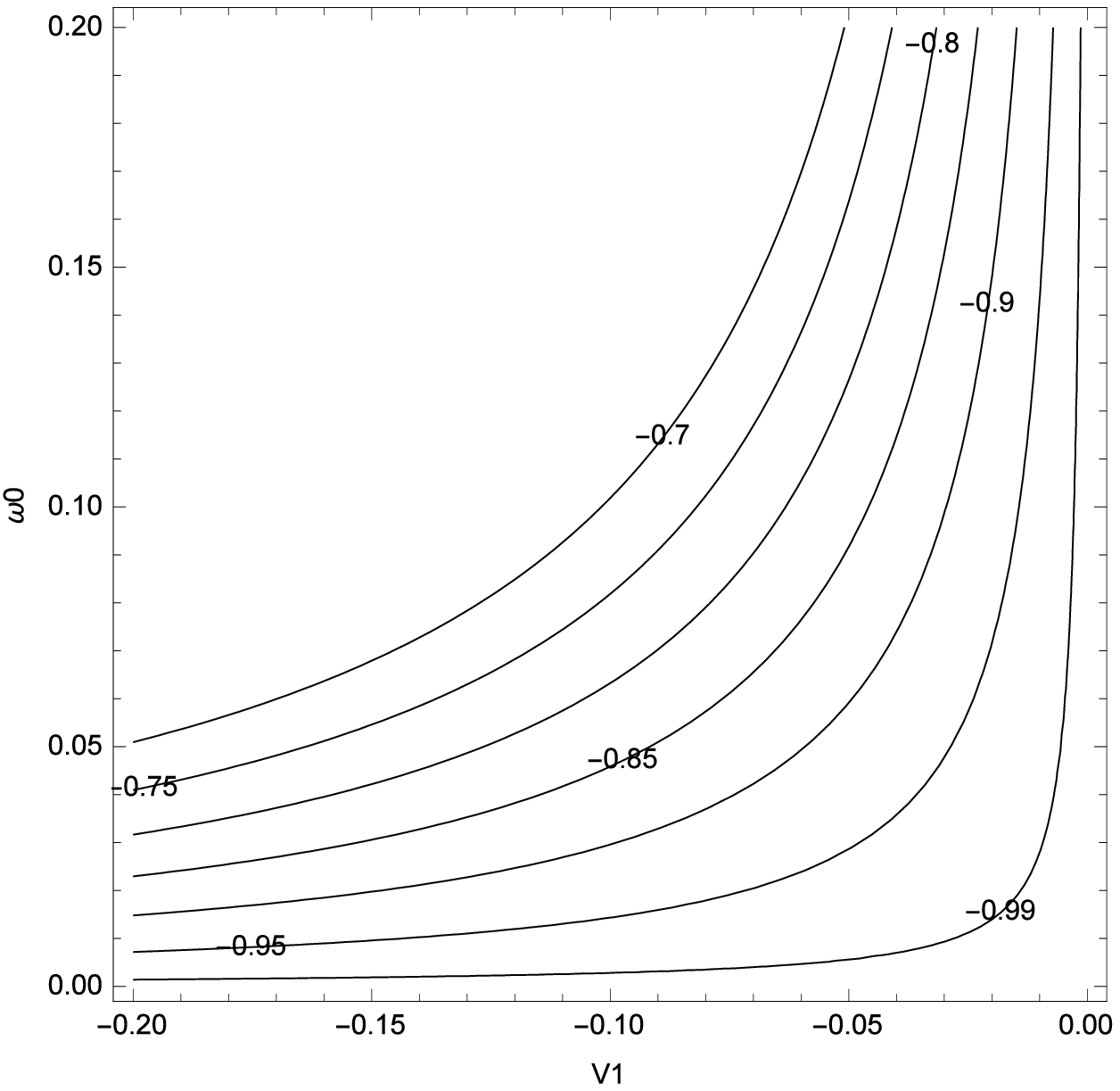}\centering
\caption{Qualitative evolution of the effective equation of state parameter
$w_{eff}$ for the Hubble function (\ref{lf.89}). Left fig. is the evolution
with respect the scale factor for~$a_{0}=1$, $\omega_{0}=0.1$ and
$V_{1}=-\frac{1}{2}10^{-3}\,\ $(solid line), $V_{1}=-10^{-2}$ (dash-dash
line), $V_{1}=-2~10^{-2}$ (dot-dot line) and $V_{1}=-3$ $10^{-2}$ (dash-dot
line). Right figure is the contour plot of $w_{eff}$ with respect to the free
parameters $V_{1}$ and $\omega_{0}$ at the scale factor $a=1$. }%
\label{plot11}%
\end{figure}

In the case where $\omega_{0}=\omega_{1}~$and $w_{m}=0$, from (\ref{lf.88}) we
find the closed-form expression of the Hubble function in terms of the scale
factor, that is,
\begin{equation}
H\left(  a\right)  =\frac{1}{12\left(  a_{0}\right)  ^{3}\left(
-V_{1}\right)  }+2\omega_{0}a^{-3}+\frac{\sqrt{1+24\left(  a_{0}\right)
^{3}\omega_{0}\left(  -V_{1}\right)  a^{-3}}}{12\left(  a_{0}\right)
^{3}\left(  -V_{1}\right)  }.\label{lf.89}%
\end{equation}
This explicitly Hubble function is used as a toy model to study the late-time
acceleration of the universe. \ In Fig. \ref{plot11}, the qualitative
evolution of the equation of state parameter for the solution (\ref{lf.89}) is
presented for various values of the free parameters.

We continue our analysis with the analysis of the critical points for the
field equations.

\section{The evolution of the flat FLRW spacetime on a phase space}

\label{SECT:II}

From equation (\ref{lf.16}) one immediately sees that the Hubble function
$H\left(  t\right)  $ can cross the value $H\left(  t\right)  =0$, from
negative to positive values, or vice-versa, since $\rho_{\phi}$ can be
negative due the friction term $3H\dot{\phi}$. Additionally, the effective
potential $V(\phi)$ is not necessarily non-negative.

We introduce the new variables \cite{Giacomini:2017yuk}:
\begin{equation}
x=\frac{\dot{\phi}}{\sqrt{(H^{2}+1)}},\quad y=\frac{V(\phi)}{6(H^{2}+1)},\quad
z=\frac{H}{\sqrt{H^{2}+1}}, \label{eq.27}%
\end{equation}
which are related through the relation
\begin{equation}
\Omega_{m}z^{2}=z(z-x)-y.
\end{equation}
where $\Omega_{m}\equiv\frac{\rho_{m}}{3H^{2}}$ is not necessarily bounded,
since $\rho_{\phi}$ and the effective potential $V(\phi)$ is not necessarily
non-negative. Although, the interval $\Omega_{m}\in\lbrack0,1]$ corresponds to
physically reasonable matter. We have assumed $w_{m}\in\lbrack-1,1]$.\newline
The evolution equations (\ref{lf.16})-(\ref{lf.19}), are written in its
autonomous form:
\begin{subequations}
\label{cosmevol}%
\begin{align}
&  x^{\prime}=-3w_{m}(z(x-z)+y)-y(\lambda(xz-2)+3)+3z^{2}%
(xz-1),\label{cosmevol1}\\
&  y^{\prime}=y\left[  6z^{3}-\lambda(x+2yz)\right]  ,\label{cosmevol2}\\
&  z^{\prime}=\left(  z^{2}-1\right)  \left(  3z^{2}-\lambda y\right)
,\label{cosmevol3}\\
&  \lambda^{\prime}=-h(\lambda)x. \label{cosmevol4}%
\end{align}
where the prime means derivative with respect to a new time variable defined
by
\end{subequations}
\[
f^{\prime}\equiv\frac{df}{d\tau}=\frac{\dot{f}}{\sqrt{H^{2}+1}},
\]
and $\lambda=-\frac{V_{,\phi}}{V}~,~h\left(  \lambda\right)  =\lambda
^{2}\left(  \frac{V_{,\phi\phi}}{\left(  V_{,\phi}\right)  ^{2}}-1\right)  .$

For the choice $z=+1$ are recovered the equations investigated in \cite{an1}.
From (\ref{cosmevol2}) it follows that the sign of $y$ (i.e., the sign of
$V(\phi)$) is invariant for the flow. Furthermore, by definition, $z$ has the
same sign with $H$, and from (\ref{cosmevol3}) we have that $z^{\prime}%
|_{z=0}=\lambda y$, which has not a definite sign. Observe that $z=\pm1$
defines two invariant sets. Additionally, $z>0$ corresponds to expanding
universe, whereas $z<0$ corresponds to contracting universes. Since the sign
of $z$ is in general not invariant for the flow, the region of the phase space
$z=0$ can be crossed which implies the existence of a transition from
contracting, to expanding universes and viceversa. Furthermore, the system
(\ref{cosmevol1})-(\ref{cosmevol3}) is form invariant under the discrete
symmetry $(x,z,\tau)\rightarrow(-x,-z,-\tau)$. So that, the fixed points
related by this symmetry have the opposite dynamical behavior. This implies
that we can investigate just the dynamics in the region $x\geq0,z\geq0$.
However, we prefer to investigate the full region of the phase space,
although, in the numerical examples we present the phase portraits for $z>0$
which corresponds to the region of cosmological interest since this leads to a
phase of late accelerated expansion. \newline
Finally, to extract some cosmological implications of the model at hand, we
use the observables
\begin{equation}
\label{Omega-q}\Omega_{\phi}=\frac{\rho_{\phi}}{3 H^{2}}, \quad q=-1-\frac
{\dot H}{H^{2}},
\end{equation}
which satisfy
\begin{equation}
\label{obs-erva-bles}\Omega_{\phi}=\frac{x z+y}{z^{2}}, \quad q=2-\frac
{\lambda y}{z^{2}},
\end{equation}
are well-defined for $z\neq0$. \newline

\subsection{Exponential potential}

\label{Sec:3.1}

Let us consider the model in which $\lambda^{\prime}$ is identically zero,
that is~$h(\lambda)=0$, so that, we obtain the effective potential
$V=V_{0}e^{-\lambda\phi}$. We study the 3D dynamical system \eqref{cosmevol1},
\eqref{cosmevol2}, \eqref{cosmevol3}, for $\lambda$ constant. In the following
we consider $w_{m}\in\left[  -1,1\right]  $. This case contains the potential
$V_{D}$ given by \eqref{lf.58} (see subsubsection \ref{sect:4.1.2}) as the
particular case $\lambda=3$.

\subsubsection{Description of the fixed points at the finite region of the
phase space.}

The (lines of) fixed points of the 3D dynamical system \eqref{cosmevol1},
\eqref{cosmevol2}, \eqref{cosmevol3}, for $\lambda$ constant are the following:

\begin{enumerate}
\item The line $A: (x,y,z)=\left(  x_{c},0,0\right)  $, that contains the
origin of coordinates. We cannot evaluate directly the expressions
\eqref{obs-erva-bles} at these points. The eigenvalues of the linearization of
\eqref{cosmevol1}, \eqref{cosmevol2}, \eqref{cosmevol3} around the fixed point
are $0,0, -\lambda x_{c}$. Thus, it is nonhyperbolic.

\item The point $B: (x,y,z)=\left(  0,0,0\right)  $. We cannot evaluate
directly the expressions \eqref{obs-erva-bles} at this point. The eigenvalues
of the linearization of \eqref{cosmevol1}, \eqref{cosmevol2},
\eqref{cosmevol3} around the fixed point are $0,0,0$. Thus, it is
nonhyperbolic.

\item The line of fixed points $C(z_{c}):(x,y,z)=\left(  0,z_{c}^{2}%
,z_{c}\right)  $, $z_{c}\in[-1,1]$, exists for $\lambda=3$. Evaluating the
expressions \eqref{obs-erva-bles} we find $\Omega_{\phi}=1, q=-1.$ Thus, this
represents a line of de-Sitter solutions. The eigenvalues of the linearization
of \eqref{cosmevol1}, \eqref{cosmevol2}, \eqref{cosmevol3} around the line of
fixed points are $0,-3 z_{c},-3 (w_{m}+1) z_{c}$. Thus, it is nonhyperbolic.

\begin{enumerate}
\item The stable manifold of $C(z_{c})$ is 2D for $0<z_{c}\leq1, w_{m}>-1$.

\item The unstable manifold of $C(z_{c})$ is 2D for $-1\leq z_{c}<0, w_{m}>-1$.
\end{enumerate}


\item $C(z_{c})$ contains the special point $D^{\pm}: (x,y,z)=\left(
0,1,\epsilon\right)  , \epsilon=\pm1$. Evaluating the expressions
\eqref{obs-erva-bles} we find $\Omega_{\phi}=1, q=-1.$ Thus, this represents
the endpoints of the previous line of de-Sitter solutions. The eigenvalues of
the linearization of \eqref{cosmevol1}, \eqref{cosmevol2}, \eqref{cosmevol3}
around the fixed point are $0,-3 \epsilon,-3 (w_{m}+1) \epsilon$. Thus, it is nonhyperbolic.

\begin{enumerate}
\item The stable manifold of $D^{+}$ is 2D for $w_{m}>-1$.

\item The unstable manifold if $D^{-}$ is 2D for $w_{m}>-1$.
\end{enumerate}


\item The points $E^{\pm}: (x,y,z)=\left(  \epsilon,0,\epsilon\right)  ,
\epsilon=\pm1$. Evaluating the expressions \eqref{obs-erva-bles} we find
$\Omega_{\phi}=1, q=2.$ So, they represents stiff solutions. The eigenvalues
of the linearization of \eqref{cosmevol1}, \eqref{cosmevol2},
\eqref{cosmevol3} around the fixed point are $6\epsilon,3 \epsilon
(1-w_{m}),\epsilon\left[  6-\lambda\right]  $.

\begin{enumerate}
\item The points are nonhyperbolic for $w_{m}=1$ or $\lambda=6$.

\item The fixed point $E^{+}$ (respectively, $E^{-}$) is a source
(respectively, a sink), for $w_{m}<1, \lambda<6$.

\item They are saddle otherwise.
\end{enumerate}


\item The points $F^{\pm}: (x,y,z)=\left(  \epsilon\frac{3 [w_{m}+1]}%
{{\lambda}},-\frac{3 [w_{m}-1]}{2 {\lambda}},\epsilon\right)  , \epsilon=\pm
1$. Evaluating the expressions \eqref{obs-erva-bles} we find $\Omega_{\phi
}=\frac{3 (w_{m}+3)}{2 \lambda},q=\frac{1}{2} (3 w_{m}+1)$. So, they represent
perfect fluid scaling solutions. The eigenvalues of the linearization of
\eqref{cosmevol1}, \eqref{cosmevol2}, \eqref{cosmevol3} around the fixed point
are $3 (w_{m}+1) \epsilon, \frac{1}{4} \epsilon\left(  3 w_{m}-3-\sqrt{3}
\sqrt{(1-w_{m}) (-16 \lambda+21 w_{m}+75)}\right)  $, \newline$\frac{1}{4}
\epsilon\left(  3 w_{m}-3+\sqrt{3} \sqrt{(1-w_{m}) (-16 \lambda+21 w_{m}%
+75)}\right)  $.

\begin{enumerate}
\item The points are nonhyperbolic for either $w_{m}=-1$, or $\lambda
=\frac{3(w_{m}+3)}{2}$, or $w_{m}=1$.

\item they are saddle otherwise.
\end{enumerate}

\item The points $G^{\pm}: (x,y,z)=\left(  \epsilon\left[  2-\frac{6}%
{{\lambda}}\right]  ,\frac{6}{{\lambda}}-1,\epsilon\right)  , \epsilon=\pm1$.
Evaluating the expressions \eqref{obs-erva-bles} we find $\Omega_{\phi}=1,
q=\lambda-4$. So, they represent accelerating solutions for $\lambda<4$. The
eigenvalues of the linearization of \eqref{cosmevol1}, \eqref{cosmevol2},
\eqref{cosmevol3} around the fixed point are $(\lambda-6) \epsilon, 2
(\lambda-3) \epsilon, \epsilon(2 \lambda-3 w_{m}-9)$.

\begin{enumerate}
\item The points are nonhyperbolic for either $\lambda=6$ or $\lambda=3$ or
$\lambda=\frac{3 (w_{m}+3)}{2}$.

\item The fixed point $G^{+}$ (respectively, $G^{-}$) is a sink (respectively,
a source) for $w_{m}>-1,\lambda<3$.

\item The fixed point $G^{+}$ (respectively, $G^{-}$) is a source
(respectively, a sink) for $w_{m}\leq1,\lambda>6$.

\item They are saddle otherwise.\begin{figure}[ptb]
\centering
\includegraphics[width=6in]{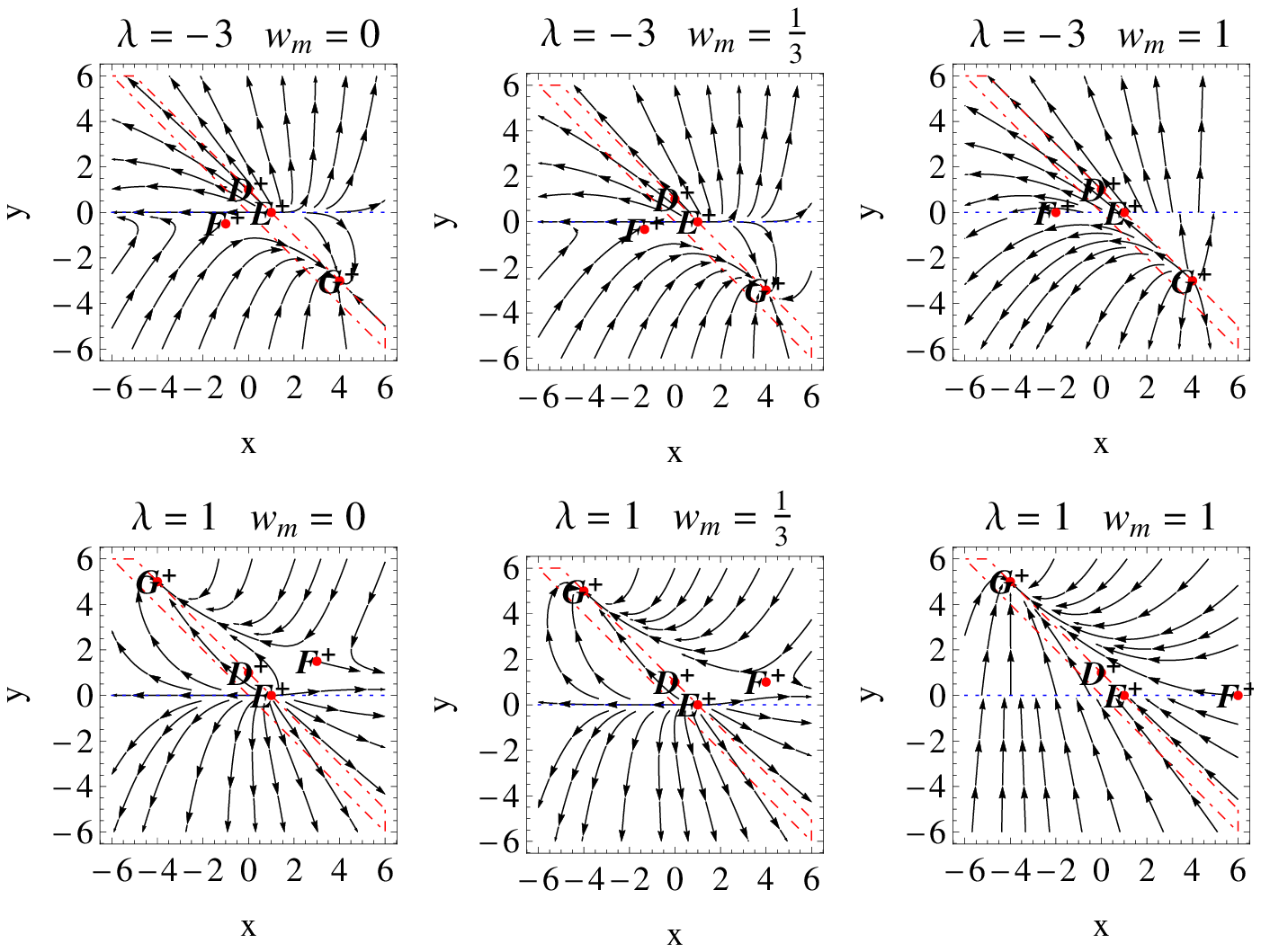} \caption{Array of phase portraits for
the restriction of the dynamical system \eqref{cosmevol} for the exponential
potential (i.e., $\lambda$ is a constant and $h\equiv0$) on the invariant set
$z=+1$ for a pressureless perfect fluid ($w_{m}=0$), a radiation fluid
($w=\frac{1}{3}$), and a stiff fluid ($w_{m}=1$) for $\lambda=-3$ and
$\lambda=1$. The dotted (blue) line denotes the invariant set $y=0$, whereas
the region enclosed by the dot-dashed (red) line corresponds to the physical
portion of the phase space.}%
\label{fig:Fig1}%
\end{figure}
\end{enumerate}
\end{enumerate}

\begin{figure}[ptb]
\centering
\includegraphics[width=6in]{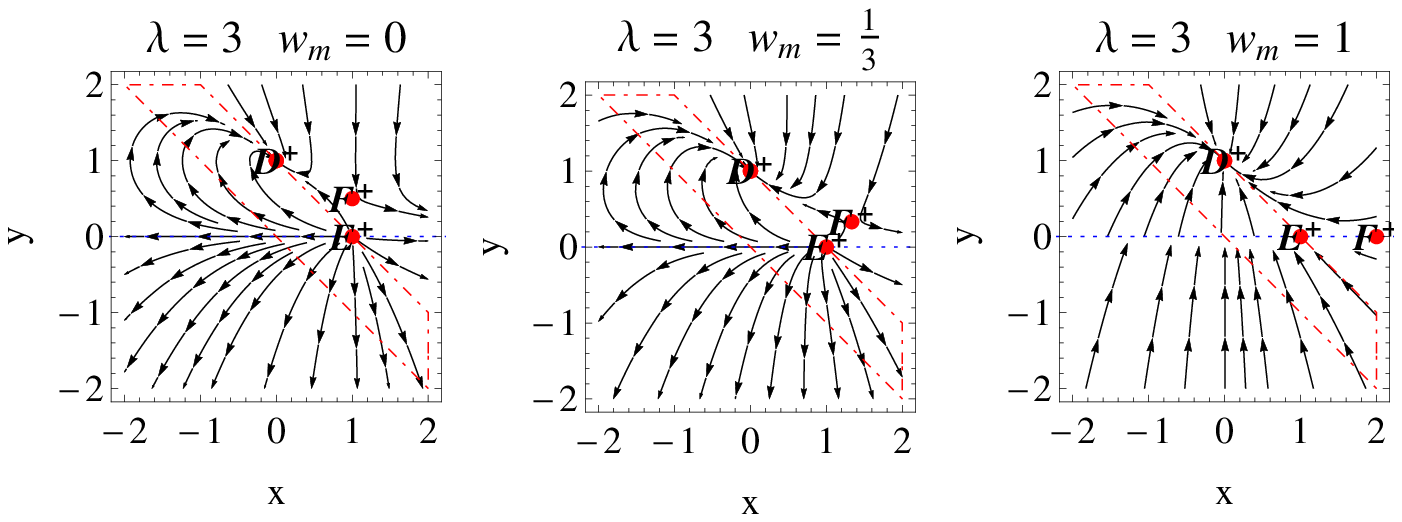} \includegraphics[width=6in]{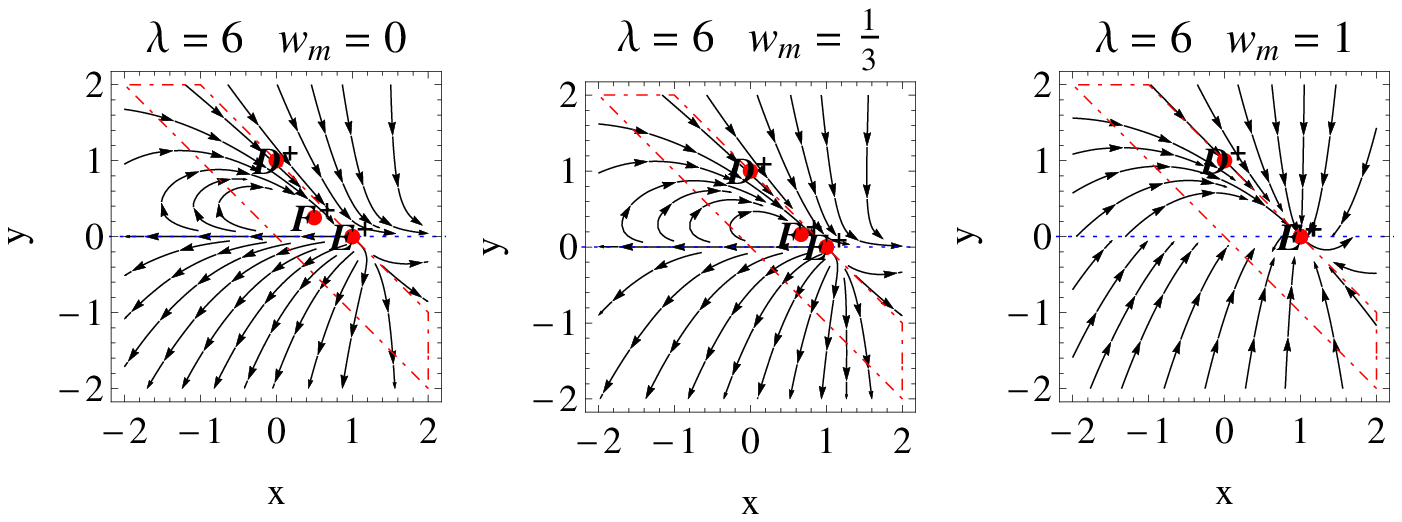}
\caption{Array of phase portraits for the restriction of the dynamical system
\eqref{cosmevol} for the exponential potential (i.e., $\lambda$ is a constant
and $h\equiv0$) on the invariant set $z=+1$ a pressureless perfect fluid
($w_{m}=0$), radiation ($w_{m}=\frac{1}{3}$), and a stiff fluid ($w_{m}=1$)
for the bifurcation parameters $\lambda=3, \lambda=6$. The dotted (blue) line
denotes the invariant set $y=0$, whereas the region enclosed by the dot-dashed
(red) line corresponds to the physical portion of the phase space. For
$\lambda=3$ the points $D^{+}$ and $G^{+}$ coincides. For $\lambda=6, w_{m}=1$
the points $D^{+}$, $F^{+}$ and $G^{+}$ coincides.}%
\label{fig:Fig2}%
\end{figure}

\subsubsection{Description of the fixed points at infinity.}

For the description of the points at infinity we introduce the variables
\begin{equation}
x=\frac{1}{\rho}\cos\psi,\quad y=\frac{1}{\rho}\sin\psi,
\end{equation}
and the time reescaling $f^{\prime}\rightarrow\rho f^{\prime}$. Defining the
new variables
\begin{equation}
X=\frac{x}{\sqrt{1+x^{2}+y^{2}}},\quad Y=\frac{y}{\sqrt{1+x^{2}+y^{2}}},
\label{compact}%
\end{equation}
we obtain that the (lines of) fixed points at infinity are:

\begin{enumerate}
\item The 2-parametric set $H(\psi,z_{c}): \left(  X,Y,z\right)  =\left(
\cos\psi, \sin\psi, z_{c}\right)  , \quad\psi\in[0, 2\pi]$, which exist for
$\lambda=0$. The eigenvalues are $0,0,0.$ The set is nonhyperbolic.

\item The points $I^{\pm}: \left(  X,Y,z\right)  =\left(  \frac{\sqrt{2}}%
{2}\epsilon,-\frac{\sqrt{2}}{2},\epsilon\right)  $, $\epsilon=\pm1$. The
eigenvalues are $\sqrt{2} \lambda\epsilon,-\frac{\sqrt{2}\lambda\epsilon}%
{2},\frac{\sqrt{2}\lambda\epsilon}{2}$. Thus, they are saddles.

\item The points $J^{\pm}: \left(  X,Y,z\right)  =\left(  -\frac{\sqrt{2}}%
{2}\epsilon, \frac{\sqrt{2}}{2}, \epsilon\right)  $, $\epsilon=\pm1$. The
eigenvalues are $-\sqrt{2}\lambda\epsilon,-\frac{\lambda\sqrt{2}\epsilon}%
{2},\frac{\lambda\sqrt{2} \epsilon}{2}$. Thus, they are saddles.

\item The lines ${}_{\pm}K(z_{c}): \left(  X,Y,z\right)  =\left(  \pm
1,0,z_{c}\right)  $, where the left subscript denotes de sign of $x$. The
eigenvalues are $0,0,\mp\lambda.$ Thus, these lines are nonhyperbolic.

\item The lines $L^{\pm}: \left(  X,Y,z\right)  =\left(  0,1,\epsilon\right)
$, $\epsilon=\pm1$. The eigenvalues are $-2 \lambda\epsilon, \lambda\epsilon,
2 \lambda\epsilon$. Thus, they are saddles.

\item The lines $M^{\pm}: \left(  X,Y,z\right)  =\left(  0, -1, \epsilon
\right)  $, $\epsilon=\pm1$. The eigenvalues are $-2 \lambda\epsilon,
-\lambda\epsilon,2 \lambda\epsilon$. Thus, they are saddles.
\end{enumerate}

We finish this section with a discussion of some numerical examples. In the
figure \ref{fig:Fig1} it is presented an array of phase portraits for the
restriction of the dynamical system \eqref{cosmevol} for the exponential
potential (i.e., $\lambda$ is a constant and $h\equiv0$) on the invariant set
$z=+1$ for a pressureless perfect fluid ($w_{m}=0$), a radiation fluid
($w=\frac{1}{3}$), and a stiff fluid ($w_{m}=1$) for $\lambda=-3$ and
$\lambda=1$. The dotted (blue) line denotes the invariant set $y=0$, whereas
the region enclosed by the dot-dashed (red) line corresponds to the physical
portion of the phase space. Furthermore, in Figure \ref{fig:Fig2} it is
presented an array of phase portraits for the restriction of the dynamical
system \eqref{cosmevol} for the exponential potential (i.e., $\lambda$ is a
constant and $h\equiv0$) on the invariant set $z=+1$ a pressureless perfect
fluid ($w_{m}=0$), radiation ($w_{m}=\frac{1}{3}$), and a stiff fluid
($w_{m}=1$) for the bifurcation parameters $\lambda=3, \lambda=6$. The dotted
(blue) line denotes the invariant set $y=0$, whereas the region enclosed by
the dot-dashed (red) line corresponds to the physical portion of the phase
space. For $\lambda=3$ the points $D^{+}$ and $G^{+}$ coincides. For
$\lambda=6, w_{m}=1$ the points $D^{+}$, $F^{+}$ and $G^{+}$ coincides.


\subsection{Beyond the Exponential Potential}

\label{Sec:3.2}

We continue our analysis with the case of a non-exponential potential in which
the dynamical system is 4D.

\subsubsection{Description of the fixed points at the finite region of the
phase space.}

The (lines of) fixed points of the 4D system \eqref{cosmevol} with finite
coordinates are:

\begin{enumerate}
\item The line $A: (x,y,z,\lambda)=\left(  x_{c},0,0,0\right)  $, $h(0)=0$. We
cannot evaluate directly the expressions \eqref{obs-erva-bles} at these
points. The eigenvalues of the linearization of \eqref{cosmevol} around the
line of fixed point are $0,0,0,-x_{c} h^{\prime}(0)$. Thus, it is
nonhyperbolic.

\item The line $B: (x,y,z,\lambda)=\left(  0,0,0,\lambda_{c}\right)  $. We
cannot evaluate directly the expressions \eqref{obs-erva-bles} at these
points. The eigenvalues of the linearization of \eqref{cosmevol} around the
line of fixed points are $0,0,0,0$. Thus, it is nonhyperbolic.

\item The line of fixed points $C(z_{c}):(x,y,z,\lambda)=\left(  0,z_{c}%
^{2},z_{c},3\right)  $, $z_{c}\in[-1,1]$. Evaluating the expressions
\eqref{obs-erva-bles} we find $\Omega_{\phi}=1, q=-1.$ Thus, this represents a
line of de-Sitter solutions. The eigenvalues of the linearization of
\eqref{cosmevol} around the line of fixed points are $0,-3 (w_{m}+1) z_{c},
-\frac{1}{2} \left(  3+\sqrt{9-8 h(3)}\right)  z_{c}, -\frac{1}{2}\left(
3-\sqrt{9-8 h(3)}\right)  z_{c}$.

\begin{enumerate}
\item The stable manifold of $C(z_{c})$ is 3D for $0<z_{c}\leq1, w_{m}>-1,
h(3)>0$.

\item The unstable manifold of $C(z_{c})$ is 3D for $-1\leq z_{c}<1, w_{m}>-1,
h(3)>0$.
\end{enumerate}


\item $C(z_{c})$ contains the special point $D^{\pm}: (x,y,z,\lambda)=\left(
0,1,\epsilon,3\right)  $, $\epsilon=\pm1$. Evaluating the expressions
\eqref{obs-erva-bles} we find $\Omega_{\phi}=1, q=-1,$ such that they are de
Sitter solutions. The eigenvalues of the linearization of \eqref{cosmevol}
around the fixed points are \newline$0,-3 (w_{m}+1) \epsilon,-\frac{1}{2}
\left(  3+\sqrt{9-8 h(3)}\right)  \epsilon, -\frac{1}{2} \left(  3-\sqrt{9-8
h(3)}\right)  \epsilon$.

\begin{enumerate}
\item The stable manifold of $D^{+}$ is 3D for $w_{m}>-1, h(3)>0$.

\item The unstable manifold of $D^{-}$ is 3D for $w_{m}>-1, h(3)>0$.
\end{enumerate}


\item The points $E^{\pm}(\hat{\lambda}): (x,y,z,\lambda)=\left(
\epsilon,0,\epsilon,\hat{\lambda}\right)  $, $\epsilon=\pm1$, and the values
$\hat{\lambda}$ satisfy $h(\hat{\lambda})=0$. Evaluating the expressions
\eqref{obs-erva-bles} we find $\Omega_{\phi}=1, q=2.$ So, they represents
stiff solutions. The eigenvalues of the linearization of \eqref{cosmevol}
around the fixed points are $6 \epsilon,3(1- w_{m}) \epsilon,\left(
6-\hat{\lambda}\right)  \epsilon,-\epsilon h^{\prime}(\hat{\lambda})$.

\begin{enumerate}
\item They are nonhyperbolic for $w_{m}=1$, or $\hat{\lambda}=6$, or
$h^{\prime}(\hat{\lambda})=0$.

\item The fixed points $E^{+}(\hat{\lambda})$ (respectively, $E^{-}%
(\hat{\lambda})$) are sources (respectively, sinks), for $w_{m}<1,
\hat{\lambda}<6, h^{\prime}(\hat{\lambda})<0$.

\item They are saddle otherwise.
\end{enumerate}


\item The points $F^{\pm}(\hat{\lambda}): (x,y,z,\lambda)=\left(
\epsilon\frac{3 [w_{m}+1]}{\hat{\lambda}},-\frac{3 [w_{m}-1]}{2 \hat{\lambda}%
},\epsilon,\hat{\lambda}\right)  $, $\epsilon=\pm1$, and the values
$\hat{\lambda}\neq0$ satisfy $h(\hat{\lambda})=0$. Evaluating the expressions
\eqref{obs-erva-bles} we find $\Omega_{\phi}=\frac{3 (w_{m}+3)}{2 \hat
{\lambda} },q=\frac{1}{2} (3 w_{m}+1)$. So, they represent perfect fluid
scaling solutions. The eigenvalues of the linearization of \eqref{cosmevol}
around the fixed points are $3 (w_{m}+1) \epsilon$,\newline$-\frac{1}{4}
\epsilon\left(  3-3 w_{m}-\sqrt{3} \sqrt{(1-w_{m}) \left(  -16 \hat{\lambda
}+21 w_{m}+75\right)  }\right)  $, \newline$-\frac{1}{4} \epsilon\left(  3-3
w_{m}+\sqrt{3} \sqrt{(1-w_{m}) \left(  -16 \hat{\lambda}+21 w_{m}+75\right)
}\right)  ,$ $-\frac{3 (w_{m}+1) \epsilon h^{\prime}(\hat{\lambda})}%
{\hat{\lambda}}$.

\begin{enumerate}
\item The points are nonhyperbolic for either $w_{m}=-1$, or $\hat{\lambda
}=\frac{3(w_{m}+3)}{2}$, or $w_{m}=1$, or $h^{\prime}(\hat{\lambda})=0$

\item They are saddle otherwise.
\end{enumerate}

\item The points $G^{\pm}(\hat{\lambda}): (x,y,z,\lambda)=\left(
\epsilon\left[  2-\frac{6}{\hat{\lambda}}\right]  ,\frac{6}{\hat{\lambda}%
}-1,\epsilon,\hat{\lambda}\right)  $, $\epsilon=\pm1$, and the values
$\hat{\lambda}\neq0$ satisfy $h(\hat{\lambda})=0$. Evaluating the expressions
\eqref{obs-erva-bles} we find $\Omega_{\phi}=1, q=\hat{\lambda} -4$. So, they
represent accelerating solutions for $\hat{\lambda}<4$. The eigenvalues of the
linearization of \eqref{cosmevol} around the fixed points are $\left(
\hat{\lambda}-6\right)  \epsilon,2 \left(  \hat{\lambda}-3\right)
\epsilon,\epsilon\left(  2 \hat{\lambda}-3 w_{m}-9\right)  , -\frac{2 \left(
\hat{\lambda}-3\right)  \epsilon h^{\prime}(\hat{\lambda})}{\hat{\lambda}}$.

\begin{enumerate}
\item The points are nonhyperbolic for either $\hat{\lambda}=6$ or
$\hat{\lambda}=3$ or $\hat{\lambda}=\frac{3 (w_{m}+3)}{2}$ or $h^{\prime}%
(\hat{\lambda})=0$.

\item The fixed point is $F^{+}$ (respectively, $F^{-}$) is a sink
(respectively, a source) for

\begin{enumerate}
\item $w_{m}>-1,\hat{\lambda}<0,h^{\prime}(\hat{\lambda})>0$ or

\item $w_{m}>-1,0<\hat{\lambda}<3,h^{\prime}(\hat{\lambda})<0$
\end{enumerate}

\item $F^{+}$ (respectively, $F^{-}$) is a source (respectively, a sink) for

\begin{enumerate}
\item $w_{m}\leq1,\hat{\lambda}>6,h^{\prime}(\hat{\lambda})<0$
\end{enumerate}

\item they are saddle otherwise.
\end{enumerate}
\end{enumerate}

\subsubsection{Description of the fixed points at infinity.}

For the description of the points when $x^{2}+y^{2}\rightarrow\infty$ we
introduce the variables
\[
x=\frac{1}{\rho}\cos\psi, \quad y=\frac{1}{\rho}\sin\psi,
\]
and the time rescaling $f^{\prime}\rightarrow\rho f^{\prime}$.

The (lines of) fixed points at infinity are:

\begin{enumerate}
\item The 2-parametric set $H(\psi,z_{c}): \left(  X,Y,z,\lambda\right)
=\left(  \cos\psi, \sin\psi, z_{c}, 0\right)  , \quad\psi\in[0, 2\pi]$, which
exist for functions $h$ satisfying , $h(0)=0$. The eigenvalues are
$0,0,0,-h^{\prime}(0) \cos\psi$.

\item The points $I^{\pm}(\hat{\lambda}): \left(  X,Y,z,\lambda\right)
=\left(  \frac{\sqrt{2}}{2}\epsilon,-\frac{\sqrt{2}}{2}, \epsilon,
\hat{\lambda}\right)  $, $\epsilon=\pm1$, and the values $\hat{\lambda}$
satisfy $h(\hat{\lambda})=0$. The eigenvalues are $-\frac{\hat{\lambda}
\epsilon}{\sqrt{2}},\frac{\hat{\lambda} \epsilon}{\sqrt{2}},\sqrt{2}
\hat{\lambda} \epsilon,-\frac{\epsilon h^{\prime}\left(  \hat{\lambda}\right)
}{\sqrt{2}}$. Thus they are saddles.

\item The points $J^{\pm}(\hat{\lambda}): \left(  X,Y,z,\lambda\right)
=\left(  -\frac{\sqrt{2}}{2}\epsilon, \frac{\sqrt{2}}{2}, \epsilon,
\hat{\lambda}\right)  $, $\epsilon=\pm1$, and the values $\hat{\lambda}$
satisfy $h(\hat{\lambda})=0$. The eigenvalues are $-\frac{\hat{\lambda}
\epsilon}{\sqrt{2}},\frac{\hat{\lambda} \epsilon}{\sqrt{2}}, -\sqrt{2}
\hat{\lambda} \epsilon,\frac{\epsilon h^{\prime}\left(  \hat{\lambda}\right)
}{\sqrt{2}}$. Thus they are saddles.

\item The lines ${}_{\pm}K(z_{c}, \hat{\lambda}): \left(  X,Y,z,\lambda
\right)  =\left(  \epsilon,0,z_{c},\hat{\lambda}\right)  $, where the left
subscript denotes de sign of $x$, and the values $\hat{\lambda}$ satisfy
$h(\hat{\lambda})=0$. The eigenvalues are $0,0,\mp\hat{\lambda} \epsilon
,\mp\epsilon h^{\prime}(\hat{\lambda})$. Thus, these lines are nonhyperbolic.

\item The lines $L^{\pm}(\lambda_{c}): \left(  X,Y,z,\lambda\right)  =\left(
0,1, \epsilon,\lambda_{c}\right)  $, $\epsilon=\pm1$. These lines of fixed
points exists independently of the functional form of $h$. The eigenvalues are
$-2 \lambda\epsilon, \lambda\epsilon, 2 \lambda\epsilon, 0$. They are normally
hyperbolic and behaves like saddles.

\item The lines $M^{\pm}(\lambda_{c}): \left(  X,Y,z,\lambda\right)  =\left(
0,-1,\epsilon, \lambda_{c}\right)  $, $\epsilon=\pm1$. These lines of fixed
points exists independently of the functional form of $h$. The eigenvalues are
$-2 \lambda\epsilon, -\lambda\epsilon, 2 \lambda\epsilon,0$. They are normally
hyperbolic and behaves like saddles.
\end{enumerate}

\subsubsection{Some specific potentials}

\label{sect:3.3} In this section we discuss some examples.

\textbf{{Example 1}}: For the potential $V(\phi)=V_{0}e^{-\sigma\phi}+V_{1},
\sigma\neq0$ and $h\equiv-\lambda(\lambda-\sigma)$. Observe that the system is
form invariant under the discrete symmetry $(x,z,\tau)\rightarrow
(-x,-z,-\tau)$. So that, the fixed points related by this symmetry have the
opposite dynamical behavior. The coordinates $(x,y,z,\lambda)$ of the fixed
points and eigenvalues of Eqs. \eqref{cosmevol} with $h\equiv-\lambda
(\lambda-\sigma)$ in the finite portion phase space with $x\geq0,z\geq0$ are
the following:

\begin{enumerate}
\item $A: (x,0,0,0)$ with eigenvalues $0,0,0,-x \sigma$. They are nonhyperbolic.

\item $A(\sigma): (x,0,0,\sigma)$ with eigenvalues $0,0,-x \sigma,x $. They
are nonhyperbolic (behaves as saddle since two eigenvalues has opposite signs).

\item $B: (0,0,0,\lambda)$ with eigenvalues $0,0,0,0 $. Thus, they are nonhyperbolic.

\item $B(\sigma): (0,0,0,\sigma)$ with eigenvalues $0,0,0,0 $. It is nonhyperbolic.

\item $C(z):\left(  0,z^{2},z,3\right)  $ with eigenvalues $0,-3 (w_{m}+1)
z,-\frac{1}{2} z \left(  \sqrt{81-24 \sigma}+3\right)  $, \newline$\frac{1}{2}
z \left(  \sqrt{81-24 \sigma}-3\right)  $.

\begin{enumerate}
\item The stable manifold of $C(z)$ is 3D for $0<z<1, w_{m}>-1, \sigma>3$.

\item The unstable manifold of $C(z)$ is 3D for $-1<z<0, w_{m}>-1, \sigma>3$.

\end{enumerate}

This line contains the points $D^{\pm}$. Due the relevance of this lines in
the cosmological setting (since they corresponds to de Sitter solutions), we
proceed forward to analyze their stability using the Center Manifold Theory.

\item $E^{+}(0): (1,0,1,0)$ with eigenvalues $6,6,3-3 w_{m},-\sigma$. Thus, it is

\begin{enumerate}
\item Nonhyperbolic for $w_{m}=1$.

\item Source for $w_{m}<1, \sigma<0$.

\item Saddle otherwise.
\end{enumerate}

\item $E^{+}(\sigma):(1,0,1,\sigma)$ with eigenvalues $\{6,3-3 w_{m}%
,6-\sigma,\sigma\}$. Thus, it is

\begin{enumerate}
\item Nonhyperbolic for $w_{m}=1$, or $\sigma=6$.

\item Source for $w_{m}<1, 0<\sigma<6$.

\item Saddle otherwise.
\end{enumerate}

\item $F^{+}(\sigma): \left(  \frac{3 (w_{m}+1)}{\sigma},-\frac{3 (w_{m}-1)}{2
\sigma}, 1,\sigma\right)  $ with eigenvalues $3 (w_{m}+1),3 (w_{m}+1)$,
\newline$\frac{1}{4} \left(  3 w_{m}-3-\sqrt{3(1-w_{m}) (21 w_{m}-16
\sigma+75)}\right)  $, \newline$\frac{1}{4} \left(  3 w_{m}-3+\sqrt{3(1-w_{m})
(21 w_{m}-16 \sigma+75)}\right)  $.

\begin{enumerate}
\item $F^{+}(\sigma)$ is nonhyperbolic for either $w_{m}=-1$, or $\sigma
=\frac{3(w_{m}+3)}{2}$, or $w_{m}=1$.

\item It is saddle otherwise.
\end{enumerate}

\item $G^{+}(\sigma): \left(  \frac{2 (\sigma-3)}{\sigma},\frac{6-\sigma
}{\sigma},1,\sigma\right)  $ with eigenvalues $\sigma-6,2 (\sigma-3),2
(\sigma-3),-3 w_{m}+2 \sigma-9 $. Thus, it is

\begin{enumerate}
\item Nonhyperbolic for $\sigma\in\left(  3,6,\frac{3}{2}(3+w_{m})\right)  $.

\item Source for $w_{m}\leq1, \sigma>6$, or $w_{m}>1, \sigma>\frac{3
(w_{m}+3)}{2}$.

\item Sink for $w_{m}\leq-1, \sigma<\frac{3 (w_{m}+3)}{2}$, or $w_{m}>-1,
\sigma<3$.

\item Saddle otherwise.
\end{enumerate}
\end{enumerate}


\textbf{{Example 2}}: Power-law potential $V\left(  \phi\right)  =\frac
{(\mu\phi)^{k}}{k}$ with $h\equiv-\frac{\lambda^{2}}{k}.$ This case contains
the potential $V_{A}$ defined by \eqref{lf.51} and discussed in subsubsection
\ref{Sect:4.1.1}, for the particular choice $k=1, V_{1}=\mu, V_{0}=0$. As
before, the system is form invariant under the discrete symmetry
$(x,z,\tau)\rightarrow(-x,-z,-\tau)$, so that, we can investigate just the
dynamics in the region $x\geq0,z\geq0$. The coordinates $(x,y,z,\lambda)$ of
the fixed points and the eigenvalues for Eqs. \eqref{cosmevol} with
$h\equiv-\frac{\lambda^{2}}{k}$ in the finite portion phase space with
$x\geq0,z\geq0$ are the following:

\begin{enumerate}
\item $A: (x,0,0,0)$ with eigenvalues $0,0,0,0 $; thus, it is nonhyperbolic.

\item $B: (0,0,0,\lambda)$ with eigenvalues $0,0,0,0 $; thus, it is nonhyperbolic.

\item $C(z): \left(  0,z^{2},z,3\right)  $ with eigenvalues\newline$0,-3
(w_{m}+1) z, -\frac{3 \left(  k+\sqrt{k (k+8)}\right)  z}{2 k},-\frac{3
\left(  k-\sqrt{k (k+8)}\right)  z}{2 k}$.

\begin{enumerate}
\item The stable manifold is 3D for $z>0, -1<w_{m}\leq1, k\leq-8$.

\item The unstable manifold is 3D for $z<0, -1<w_{m}\leq1, k\leq-8$.

\end{enumerate}

This line contains the points $D^{\pm}$.

\item $E^{+}(0): (1,0,1,0)$ with $\{6,6,0,3-3 w_{m}\}$; thus, it is nonhyperbolic.
\end{enumerate}

Since all the fixed points are nonhyperbolic we rely on numerical inspection.
However, for the line of de Sitter solutions $C(z)$ we implement the Center
Manifold computation.


\textbf{{Example 3}}: Hyperbolic Potential $V(\phi)=V_{0}(\cosh(\xi\phi)-1),
\xi\neq0$, and $h\equiv-\frac{1}{2}\left(  \lambda^{2}-\xi^{2}\right)  $. As
before, the system is form invariant under the discrete symmetry
$(x,z,\tau)\rightarrow(-x,-z,-\tau)$, so that, we can investigate just the
dynamics in the region $x\geq0,z\geq0$. The coordinates $(x,y,z,\lambda)$ of
the fixed points and eigenvalues for Eqs. \eqref{cosmevol} with $h\equiv
-\frac{1}{2}\left(  \lambda^{2}-\xi^{2}\right)  $ in the finite portion phase
space with $x\geq0,z\geq0$ are given by:

\begin{enumerate}
\item $A(-\xi): (x,0,0,-\xi)$ with eigenvalues $0,0,-x \xi,x \xi$. They are
nonhyperbolic (behaves a saddles).

\item $A(\xi): (x,0,0,\xi)$ with eigenvalues $0,0,-x \xi,x \xi$. They are
nonhyperbolic (behaves a saddles).

\item $B: (0,0,0,\lambda)$ with eigenvalues $0,0,0,0 $. They are nonhyperbolic.

\item $B(-\xi): (0,0,0,-\xi)$ with eigenvalues $0,0,0,0 $. It is nonhyperbolic.

\item $B(\xi): (0,0,0,\xi)$ with eigenvalues $0,0,0,0 $. It is nonhyperbolic.

\item $C(z): \left(  0,z^{2},z,3\right)  $ with eigenvalues\newline$0,-3
(w_{m}+1) z,\frac{z}{2} \left(  -3-\sqrt{ 45-4 \xi^{2}}\right)  ,\frac{z}{2}
\left(  -3+\sqrt{45-4 \xi^{2}}\right)  $.

\begin{enumerate}
\item The stable manifold is 3D for $z>0, -1<w_{m}\leq1, -\frac{3 \sqrt{5}}%
{2}\leq\xi<-3$ or $z>0, -1<w_{m}\leq1, 3<\xi\leq\frac{3 \sqrt{5}}{2}$.

\item The unstable manifold is 3D for $z<0, -1<w_{m}\leq1, -\frac{3 \sqrt{5}%
}{2}\leq\xi<-3$ or $z<0, -1<w_{m}\leq1, 3<\xi\leq\frac{3 \sqrt{5}}{2}$.

\end{enumerate}

This line contains the points $D^{\pm}$.

\item $E^{+}(-\xi): (1,0,1,-\xi)$ with eigenvalues $6,3-3 w_{m},-\xi,\xi+6 $.
It is

\begin{enumerate}
\item Nonhyperbolic for $w_{m}=1$, or $\xi=-6$.

\item source for $-6<\xi<0, w_{m}<1$.

\item saddle otherwise.
\end{enumerate}

\item $E^{+}(\xi): (1,0,1,\xi)$ with eigenvalues $6,3-3 w_{m},6-\xi,\xi$. It is

\begin{enumerate}
\item Nonhyperbolic for $w_{m}=1$, or $\xi=6$.

\item source for $0<\xi<6, w_{m}<1$.

\item saddle otherwise.
\end{enumerate}

\item $F^{+}(-\xi): \left(  -\frac{3 (w_{m}+1)}{\xi},\frac{3 (w_{m}-1)}{2 \xi
},1,-\xi\right)  $ with eigenvalues \newline$3 (w_{m}+1),3 (w_{m}+1),\frac{3
(w_{m}-1) -\sqrt{3} \sqrt{-(w_{m}-1) (21 w_{m}+16 \xi+75)}}{4},\frac{3
(w_{m}-1) +\sqrt{3} \sqrt{-(w_{m}-1) (21 w_{m}+16 \xi+75)}}{4} $.

\begin{enumerate}
\item $F^{+}(-\xi)$ is nonhyperbolic for either $w_{m}=-1$, or $\xi
=-\frac{3(w_{m}+3)}{2}$, or $w_{m}=1$.

\item It is saddle otherwise.
\end{enumerate}

\item $F^{+}(\xi): \left(  \frac{3 (w_{m}+1)}{\xi},-\frac{3 (w_{m}-1)}{2 \xi
},1,\xi\right)  $ with eigenvalues\newline$3 (w_{m}+1),3 (w_{m}+1),\frac{3
(w_{m}-1)-\sqrt{3} \sqrt{-(w_{m}-1) (21 w_{m}-16 \xi+75)}}{4}, \frac{3
(w_{m}-1) +\sqrt{3} \sqrt{-(w_{m}-1) (21 w_{m}-16 \xi+75)}}{4} $.

\begin{enumerate}
\item $F^{+}(\xi)$ is nonhyperbolic for either $w_{m}=-1$, or $\xi
=\frac{3(w_{m}+3)}{2}$, or $w_{m}=1$.

\item It is saddle otherwise.
\end{enumerate}

\item $G^{+}(-\xi): \left(  \frac{2 (\xi+3)}{\xi},\frac{-\xi-6}{\xi}%
,1,-\xi\right)  $ with eigenvalues\newline$-2 (\xi+3),-2 (\xi+3),-3 w_{m}-2
\xi-9,-\xi-6 $. It is

\begin{enumerate}
\item Nonhyperbolic for $\xi\in\left(  -3,-6, -\frac{3}{2}(3+w_{m})\right)  $.

\item Source for $w_{m}\leq1, \xi<-6$ or $w_{m}>1, \xi<-\frac{3(w_{m}+3)}{2}$.

\item Sink for $w_{m}\leq-1, \xi>-\frac{3 (w_{m}+3)}{2}$ or $w_{m}>-1, \xi>-3$.

\item Saddle otherwise.
\end{enumerate}

\item $G^{+}(\xi): \left(  \frac{2 (\xi-3)}{\xi},\frac{6-\xi}{\xi}%
,1,\xi\right)  $ with eigenvalues $\xi-6,2 (\xi-3),2 (\xi-3),-3 w_{m}+2 \xi-9
$. It is

\begin{enumerate}
\item Nonhyperbolic for $\xi\in\left(  3,6, \frac{3}{2}(3+w_{m})\right)  $.

\item Source for $w_{m}\leq1, \xi> 6$ or $w_{m}>1, \xi>\frac{3(w_{m}+3)}{2}$.

\item Sink for $w_{m}\leq-1, \xi<\frac{3 (w_{m}+3)}{2}$ or $w_{m}>-1, \xi<3$.

\item Saddle otherwise.
\end{enumerate}
\end{enumerate}


\subsection{Critical points for potentials supported by Cartan symmetries}

Finally, we discuss some models that were introduced by the Cartan symmetries
in Section \ref{cartan2}. \newline{\textbf{Example 4:}} For the potential
$V_{B}\left(  \phi\right)  $ we calculate $h_{B}\equiv-\left(  \lambda
-3(w_{m}+1)\right)  (\lambda-6w_{m})~$. Due to the existence of the discrete
symmetry $(x,z,\tau)\rightarrow(-x,-z,-\tau)$, the fixed points related by
this symmetry have the opposite dynamical behavior. The coordinates
$(x,y,z,\lambda)$ of the fixed points and eigenvalues for Eqs.
\eqref{cosmevol} with $h\equiv-(\lambda-6w_{m})\left(  \lambda-3(w_{m}%
+1)\right)  $ in the finite portion phase space with $x\geq0,z\geq0$ are the following.

\begin{enumerate}
\item $A_{1}: (x,0,0,6 w_{m})$ with eigenvalues $0,0,3 (w_{m}-1) x,-6 w_{m} x
$.

\begin{enumerate}
\item The stable manifold of $A_{1}$ is 2D for $0<w_{m}<1, x>0$.

\item The unstable manifold of $A_{1}$ is 2D for $0<w_{m}<1, x<0$.
\end{enumerate}

\item $A_{2}: (x,0,0,3 (w_{m}+1))$ with eigenvalues $0,0,-3 (w_{m}-1) x,-3
(w_{m}+1) x $. The nonzero eigenvalues has different signs for $-1<w_{m}<1,
x\neq0$. Thus, it behaves as a saddle.

\item $B: (0,0,0,\lambda)$ with eigenvalues $0,0,0,0 $. They are nonhyperbolic.

\item $B_{1}: (0,0,0,6 w_{m})$ with eigenvalues $0,0,0,0 $. It is nonhyperbolic.

\item $B_{2}: (0,0,0,3 (w_{m}+1))$ with eigenvalues $0,0,0,0 $. It is nonhyperbolic.

\item $C(z): \left(  0,z^{2},z,3\right)  $ with eigenvalues $0,-6 w_{m} z,-3
(w_{m}+1) z,3 (2 w_{m}-1)z $.

\begin{enumerate}
\item Its stable manifold is 3D for $0<w_{m}<\frac{1}{2}, z>0$.

\item Its unstable manifold is 3D for $0<w_{m}<\frac{1}{2}, z<0$.
\end{enumerate}

This curve contains the points $D^{\pm}$.

\item $E^{+}_{1}: (1,0,1,6 w_{m})$ with eigenvalues $6,3-3 w_{m},6-6 w_{m},3
(w_{m}-1) $.

\begin{enumerate}
\item It is nonhyperbolic for $w_{m}=1$.

\item It is a saddle otherwise.
\end{enumerate}

\item $E^{+}_{2}: (1,0,1,3 (w_{m}+1))$ with eigenvalues $6,3-3 w_{m},3-3
w_{m},3-3 w_{m} $.

\begin{enumerate}
\item It is nonhyperbolic for $w_{m}=1$.

\item It is a source for $w_{m}<1$.
\end{enumerate}

\item $F^{+}_{1}: \left(  \frac{w_{m}+1}{2 w_{m}}, \frac{1-w_{m}}{4 w_{m}}, 1,
6 w_{m}\right)  , w_{m}\neq0$, with eigenvalues $\frac{3 \left(  w_{m}%
^{2}-1\right)  }{2 w_{m}}, 3 (w_{m}+1)$, \newline$3-3w_{m}, \frac{9 (w_{m}%
-1)}{2} $.

\begin{enumerate}
\item It is nonhyperbolic for $w_{m} \in\{-1,1\}$.

\item Saddle otherwise.
\end{enumerate}

\item $F^{+}_{2}: \left(  1,\frac{1-w_{m}}{2 (w_{m}+1)},1,3 (w_{m}+1)\right)
, w_{m}\neq-1$, with eigenvalues\newline$3-3w_{m},-\frac{3}{2} (w_{m}-1),3
(w_{m}-1),3 (w_{m}+1) $.

\begin{enumerate}
\item It is nonhyperbolic for $w_{m} =1$.

\item Saddle otherwise.
\end{enumerate}

\item $G^{+}_{1}: \left(  \frac{2 w_{m}-1}{w_{m}},\frac{1-w_{m}}{w_{m}},1,6
w_{m}\right)  , w_{m}\neq0$, with eigenvalues $6 w_{m}-9+\frac{3}{w_{m}}$,
\newline$12 w_{m}-6,6 (w_{m}-1),9 (w_{m}-1) $.

\begin{enumerate}
\item Nonhyperbolic for $w_{m}\in\left(  \frac{1}{2},1\right)  $

\item It is a sink for $-1\leq w_{m}<0$.

\item It is a saddle otherwise.
\end{enumerate}

\item $G^{+}_{2}: \left(  \frac{2 w_{m}}{w_{m}+1}, \frac{1-w_{m}}{w_{m}+1}, 1,
3 (w_{m}+1)\right)  , w_{m} \neq-1$, with eigenvalues $-\frac{6 (w_{m}-1)
w_{m}}{w_{m}+1}, 6 w_{m}$, \newline$3 (w_{m}-1), 3(w_{m}-1) $.

\begin{enumerate}
\item Nonhyperbolic for $w_{m}\in\left(  0,1\right)  $

\item It is a sink for $-1< w_{m}<0$.

\item It is a saddle otherwise.
\end{enumerate}
\end{enumerate}

We have used subscripts to distinguish each particular member of a class,
instead to specify $\hat{\lambda}$, to avoid a cumbersome notation. The
subscript $1$ means evaluation at $\hat{\lambda}=6w_{m}$, whereas, the
subscript $2$ means evaluation at $\hat{\lambda}=3 (w_{m}+1)$.\newline
\newline{\textbf{Example 5:}} For the potential $V_{C}\left(  \phi\right)  $
we calculate $h_{C}\equiv-\frac{1}{2}\left(  \lambda-3(w_{m}+1)\right)
\left(  2\lambda-3(3+w_{m})\right)  ~$. Due to the existence of the discrete
symmetry $(x,z,\tau)\rightarrow(-x,-z,-\tau)$, the fixed points related by
this symmetry have the opposite dynamical behavior. The coordinates
$(x,y,z,\lambda)$ of the fixed points and eigenvalues for Eqs.
\eqref{cosmevol} with $h\equiv-\frac{1}{2}\left(  \lambda-3(w_{m}+1)\right)
\left(  2\lambda-3(3+w_{m})\right)  $ in the finite portion phase space with
$x\geq0,z\geq0$ are the following.

\begin{enumerate}
\item $A_{1}: \left(  x,0,0,\frac{3 (w_{m}+3)}{2}\right)  $ with eigenvalues
$0,0,-\frac{3}{2} (w_{m}-1) x,-\frac{3}{2} (w_{m}+3) x $. They are
nonhyperbolic (behaves as saddles for $x\neq0$).

\item $A_{2}: (x,0,0,3 (w_{m}+1))$ with eigenvalues $0,0,\frac{3}{2} (w_{m}-1)
x, -3 (w_{m}+1) x $. They are nonhyperbolic (behaves as saddles for $x\neq0$).

\item $B: (0,0,0,\lambda)$ with eigenvalues $0,0,0,0 $. They are nonhyperbolic.

\item $B_{1}: \left(  0,0,0,\frac{3 (w_{m}+3)}{2}\right)  $ with eigenvalues
$0,0,0,0 $. It is nonhyperbolic.

\item $B_{2}: (0,0,0,3 (w_{m}+1))$ with eigenvalues $0,0,0,0 $. It is nonhyperbolic.

\item $C(z): \left(  0,z^{2},z,3\right)  $ with eigenvalues $0,3 w_{m} z,-3
(w_{m}+1) z,-3 (w_{m}+1) z $.

\begin{enumerate}
\item Its stable manifold is 3D for $-1<w_{m}<0, z>0$.

\item Its unstable manifold is 3D for $-1<w_{m}<0, z<0$.
\end{enumerate}

This curve contains the points $D^{\pm}$.

\item $E^{+}_{1}: \left(  1,0,1,\frac{3 (w_{m}+3)}{2}\right)  $ with
eigenvalues $6,-3(w_{m}-1),-\frac{3}{2} (w_{m}-1),-\frac{3}{2} (w_{m}-1) $.

\begin{enumerate}
\item It is nonhyperbolic for $w_{m}=1$.

\item It is a source for $w_{m}<1$.
\end{enumerate}

\item $E^{+}_{2}: (1,0,1,3 (w_{m}+1))$ with eigenvalues $6,-3(w_{m}-1),
-3(w_{m}-1), \frac{3 (w_{m}-1)}{2} $.

\begin{enumerate}
\item It is nonhyperbolic for $w_{m}=1$.

\item It is a saddle otherwise.
\end{enumerate}

\item $F^{+}_{1}: \left(  \frac{2 (w_{m}+1)}{w_{m}+3},\frac{1-w_{m}}{w_{m}%
+3},1,\frac{3 (w_{m}+3)}{2}\right)  $ with eigenvalues $0,\frac{3-3 w_{m}^{2}%
}{w_{m}+3},3 (w_{m}+1),\frac{3 (w_{m}-1)}{2} $. It is nonhyperbolic. The zero
eigenvalue appears due to the bifurcation value $\hat{\lambda}$, where
$F^{+}_{1}$ and $G^{+}_{1}$ coincide. It behaves a saddle (at least two
eigenvalues are of different sign).

\item $F^{+}_{2}: \left(  1,\frac{1-w_{m}}{2 (w_{m}+1)},1,3 (w_{m}+1)\right)
, w_{m}\neq-1$, with eigenvalues\newline$-\frac{3}{2} (w_{m}-1),\frac{3
(w_{m}-1)}{2},3 (w_{m}-1),3 (w_{m}+1) $.

\begin{enumerate}
\item It is nonhyperbolic for $w_{m}=1$.

\item It is a saddle otherwise.
\end{enumerate}

\item $G^{+}_{1}$ merges with $F^{+}_{1}$. Thus, it behaves as saddle.

\item $G^{+}_{2}: \left(  \frac{2 w_{m}}{w_{m}+1},\frac{1-w_{m}}{w_{m}+1},1,3
(w_{m}+1)\right)  $ with eigenvalues\newline$\frac{3 (w_{m}-1) w_{m}}{w_{m}%
+1},6 w_{m},3 (w_{m}-1),3(w_{m}-1) $.

\begin{enumerate}
\item Nonhyperbolic for $w_{m}\in\left(  0,1\right)  $

\item It is a saddle otherwise.
\end{enumerate}
\end{enumerate}

We have used subscripts to distinguish each particular member of a class,
instead to specify $\hat{\lambda}$, to avoid a cumbersome notation. The
subscript $1$ means evaluation at $\hat{\lambda}=\frac{3 (w_{m}+3)}{2}$,
whereas, the subscript $2$ means evaluation at $\hat{\lambda}=3 (w_{m}%
+1)$.\newline

\section{Toy model and Supernova data}

\label{cartan3}

Consider now the Hubble parameter
\begin{equation}
E(a)=\frac{H\left(  a\right)  }{H_{0}}=\Omega_{\Lambda0}\left(  1+\sqrt
{1+\frac{\Omega_{s0}}{\Omega_{\Lambda0}}a^{-3}}\right)  +\Omega_{s0}a^{-3},
\label{lf.90}%
\end{equation}
where if we compare it with (\ref{lf.89}) it follows that $\Omega_{\Lambda
0}=\left(  12\left(  a_{0}\right)  ^{3}\left(  -V_{1}\right)  H_{0}\right)
^{-1}$ and \thinspace$\Omega_{s0}=2\omega_{0}H_{0}^{-1}$. It is clear that
from that Hubble function, except from the cosmological constant term and the
stiff fluid, there is also a term which provides a dark energy component. This
is not the first time that this noncanonical scalar field provide dust terms
in the cosmological solution. It has been observed before in \cite{an1,anbas}.

Moreover, from the constraint $H\left(  a\rightarrow1\right)  =H_{0}$, we find
the algebraic relation between the two free parameters $\Omega_{\Lambda0}$ and
$\Omega_{s0}$,%
\begin{equation}
\Omega_{\Lambda0}=\frac{1-\Omega_{s0}}{2-\Omega_{s0}}. \label{lf.91}%
\end{equation}
which is used to reduce the free parameters of the model. It is interesting to
mention that the current model contains the same number of free parameters
with that of the concordance $\Lambda$CDM model.

We continue by constraining the Hubble function (\ref{lf.90}) with some of the
cosmological data. In particular we perform a joint likelihood analysis in
order to constraint the one free parameter, $\Omega_{s0}$, by using the SNIa
data of the Union 2.1 collaboration \cite{Suzuki}.

The likelihood function is determined to be $\mathcal{L}\mathcal{=}%
e^{-\chi_{A}^{2}/2}~$; that is, $\chi^{2}=\chi_{SNIa}^{2}$ and the Likelihood
function is maximized for the minimum parameter of $\chi^{2}$. The Union 2.1
data set provides us with 580 SNIa distance modulus at observed redshift
\cite{Suzuki} with observed redshift in the range $z_{i}\in$ $\left[
0.015,1.414\right]  $. The chi-square parameter for the diagonal covariant
matrix is given by the expression
\begin{equation}
\chi_{SNIa}^{2}(\mathbf{\epsilon})=\sum\limits_{i=1}^{N_{SNIa}}\left(
\frac{\mu_{obs}\left(  z_{i}\right)  -\mu_{th}\left(  z_{i},\mathbf{\epsilon
}\right)  }{\sigma_{i}}\right)  ^{2} \label{lf.93}%
\end{equation}
where $\mathbf{\epsilon}\equiv\{H_{0},p^{1},p^{2},...\}$ denotes the
statistical vector that contains the free parameters of the model,
$N_{SNIa}=580$, $z_{i}$ is the observed redshift, $\mu_{obs}$ is the observed
distance modulus and $\mu_{th}$ is the theoretical distance modulus which is
given by
\begin{equation}
\mu=m-M=5\log d_{L}+25=5\log D_{L}+\mu_{0}, \label{MMOD}%
\end{equation}
where
\begin{equation}
d_{L}(\mathbf{\epsilon},z)=\frac{c}{H_{0}}D_{L}(p^{j},z)=\frac{c}{H_{0}}%
\int_{0}^{z}\frac{dx}{E(x,p^{j})}%
\end{equation}
and $\mu_{0}=42.384-5\log h$ with $h=H_{0}/100$. Including the second equality
of Eq.(\ref{MMOD}) into Eq.({\ref{lf.93}) we arrive at
\begin{equation}
\chi_{SNIa}^{2}(\mathbf{\epsilon})=A-2B\mu_{0}+\Gamma\mu_{0}^{2},
\label{MMOD1}%
\end{equation}
where
\begin{equation}
A(p^{j})=\sum\limits_{i=1}^{N_{SNIa}}\left(  \frac{\mu_{obs}\left(
z_{i}\right)  -5\log D_{L}\left(  z_{i},p^{j}\right)  }{\sigma_{i}}\right)
^{2} \label{AM1}%
\end{equation}%
\begin{equation}
B(p^{j})=\sum\limits_{i=1}^{N_{SNIa}}\frac{\mu_{obs}\left(  z_{i}\right)
-5\log D_{L}\left(  z_{i},p^{j}\right)  }{\sigma_{i}^{2}} \label{BM1}%
\end{equation}%
\begin{equation}
\Gamma=\sum\limits_{i=1}^{N_{SNIa}}\frac{1}{\sigma_{i}^{2}}\;. \label{CM1}%
\end{equation}
Clearly, for $\mu_{0}=B/\Gamma$, (\ref{MMOD1}) has a minimum at
\[
{\tilde{\chi}}_{SNIa}^{2}(p^{j})=A(p^{j})-\frac{\left(  B(p^{j})\right)  ^{2}%
}{\Gamma}.
\]
The latter implies that instead of using $\chi_{SNIa}^{2}(\mathbf{\epsilon})$
we now minimize ${\tilde{\chi}}_{SNIa}^{2}(p^{j})$ which is independent of
$\mu_{0}$ and hence of the value of the Hubble constant. Therefore, for the
current model we have only one free parameter, namely $p^{1}=\Omega_{s0}$. The
reader may find more details regarding the aforementioned statistical
procedure in \cite{NesPeriv}. }

We compare the model (\ref{lf.90}) with that of the $\Lambda$-cosmology~whose
Hubble function is
\begin{equation}
\frac{H_{\Lambda}\left(  a\right)  }{H_{0}}=\sqrt{\left(  1-\Omega
_{m0}\right)  +\Omega_{m0}a^{-3}}. \label{lf.95}%
\end{equation}
In this case the free parameter of the model is $p^{1}=\Omega_{m0}$.

From the SNIa data we found that $\left(  \min\chi^{2}\right)  =562.77$ while
the best fit value is $\Omega_{s0}=0.0835_{-0.055}^{+0.065}.$ With the same
data for the $\Lambda$-cosmology we find that $\Omega_{m0}=0.29$ with $\left(
\min\chi^{2}\right)  ^{\Lambda}=561.73$.

The two models have the same number of degrees of freedom and the difference
of the minimum $\chi^{2}$ is approximately one. Therefore according to the
Akaike information criterion \cite{Akaike1974,sugu} the two models fit the
Supernova data with the same way.

Of course, model (\ref{lf.90}) has been used as a toy model in order to show
that the model we proposed and the solutions which result provide parameters
which allow it to fit the cosmological observations. Further extended analysis
is required, which however is beyond the scope of the present study.

\section{Evolution of the observables}

\label{Sect:7}

Following the reference \cite{WE}, we choose $t=0$ corresponding to the
initial singularity, and denote $t_{0}$ as the age of the universe. The
current value $H_{0}$ of the Hubble scalar is called the Hubble constant. For
these quantities we have observable bounds. Now, we introduce the
dimensionless parameters
\begin{equation}
\alpha=tH,\beta=\frac{\dot{\phi}}{H}.\label{alpha}%
\end{equation}
The present value of $\alpha$, denoted by $\alpha_{0}=t_{0}H_{0}$ is referred
as the age parameter and it is a well-defined function in state space
\cite{WE}. In an ever expanding model, where $a=a_{0}e^{N}$, the numbers of
e-foldings $N$ assume all real values, thus we can study the dynamical system
\begin{subequations}
\label{COSMOEVOL2}%
\begin{align}
&  \frac{d{\Omega_{\phi}}}{dN}=(\Omega_{\phi}-1)(2q-3w_{m}-1),\\
&  \frac{d{\beta}}{dN}=(q(\beta-2)-3(w_{m}+1)\Omega_{\phi}+3w_{m}+4\beta+1),\\
&  \frac{d{\lambda}}{dN}=-\beta h(\lambda),\\
&  q=2+\lambda(\beta-\Omega_{\phi}),\label{eq7-2-d}%
\end{align}
and the decoupled equation
\end{subequations}
\begin{equation}
\frac{d{\alpha}}{dN}=1-(1+q)\alpha.
\end{equation}
The latter algebraic-differential system is exactly the system
(\ref{cosmevol1})-(\ref{cosmevol4}) but in different variables.

Let us denote by $y$ the vector $\left(  \Omega_{\phi},\beta,\lambda\right)
$. We have seen that $q$ is a function of the phase space as defined by
\eqref{eq7-2-d}. Hence, at a fixed point $y^{\star}$ of the DE
\eqref{COSMOEVOL2}, $q$ is a constant, i.e., $q(y^{\star})$, (the particular
values of $q$ are summarized in the Appendix \ref{appendixB}). Given an
initial point $\mathbf{y}_{0}$ - which represents our universe in the present
time, let denoted by $\mathbf{y}=\Phi_{N}(\mathbf{y}_{0})$ the orbit through
$\mathbf{y}_{0}$ with $\Phi_{0}(\mathbf{y}_{0})=\mathbf{y}_{0}$, and by
\begin{equation}
\tilde{q}(N)=q(\Phi_{N}(\mathbf{y}_{0})),
\end{equation}
the deceleration parameter along the orbit so that $\tilde{q}(0)=q(\mathbf{y}%
_{0})$.

Then, are deduced the expressions \cite{WE}:
\begin{subequations}
\begin{align}
&  H(N)=H_{0} \exp\left[  -\int_{0}^{N} \left\{  1+\tilde{q}(\mu)\right\}  d
\mu\right]  ,\;\text{for all}\; N\in\mathbb{R}.\\
&  t_{0}=\int_{-\infty}^{0} \frac{1}{H(N)} dN,\\
&  t_{0} H_{0}=\int_{-\infty}^{0} \exp\left[  \int_{0}^{N} \left\{
1+\tilde{q}(\mu)\right\}  d \mu\right]  dN
\end{align}
where $H_{0}$ is a freely specifiable. This arbitrariness implies that each
non-singular orbit corresponds to a 1-parameter family of physical universes,
which are conformally related by a constant rescaling of the metric.
$t_{0}=t(0)$, denotes the value of $t$ at $\mathbf{y}_{0}$. The last formula
implies that $\alpha_{0}=t_{0} H_{0}$ is uniquely determined by the specified
initial point $\mathbf{y}_{0}$ on the phase space, such that $\alpha_{0}=t_{0}
H_{0}$ is a well-defined function on state space. Furthermore, the constraints
$0.87 <\alpha(\mathbf{y}_{0}) <1.68$, $0.1\lesssim\Omega_{0}\lesssim0.3$
\cite{WE}, where $\Omega_{0}=1-\Omega_{\phi}(\mathbf{y}_{0})$, will restrict
the location of the present state of the universe, $\mathbf{y}_{0}$, in state space.


\begin{figure}[ptb]
\centering
\includegraphics[width=6in]{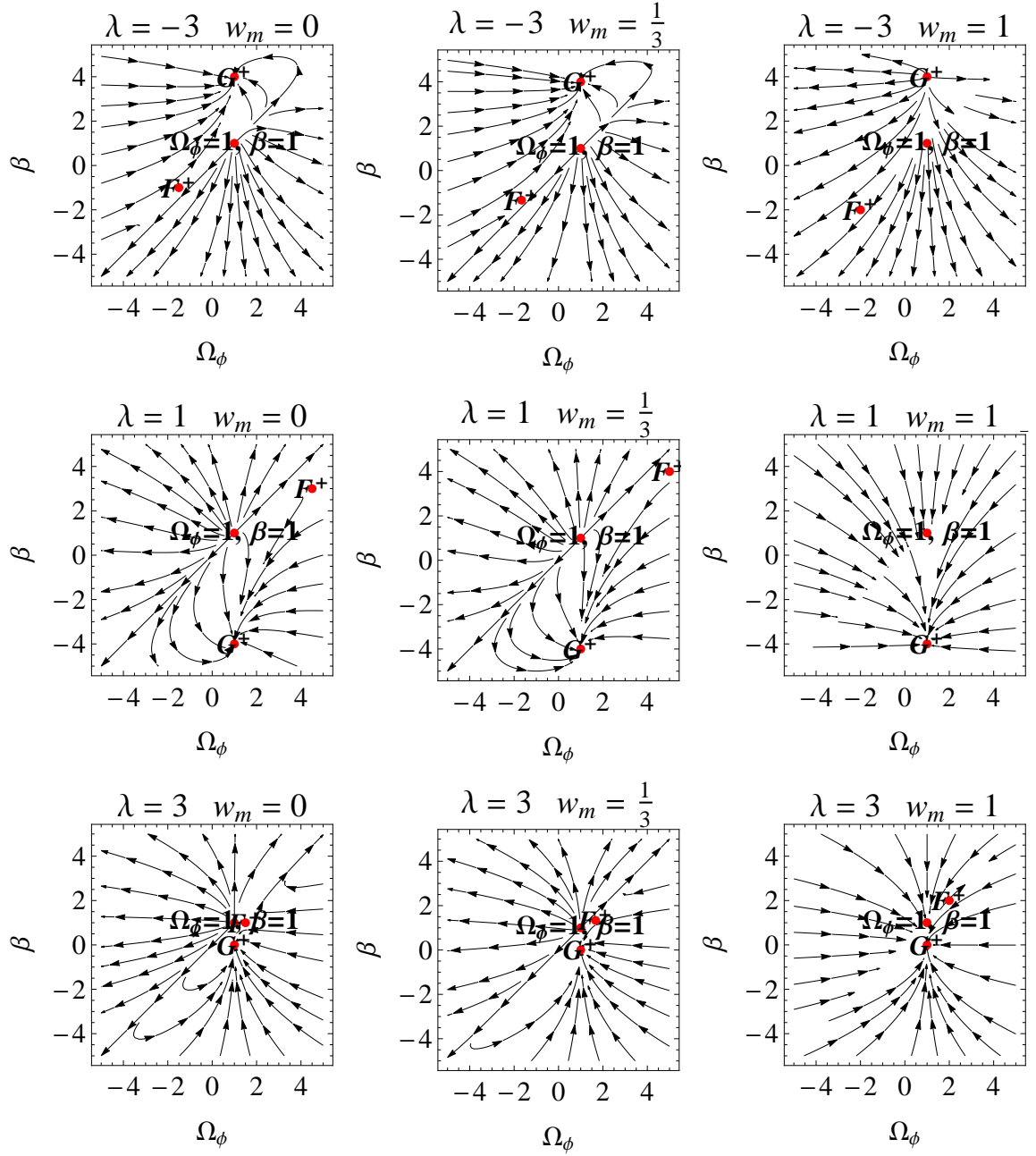}\caption{Evolution of the system
\eqref{COSMOEVOL2} for an exponential potential for some choices of the
parameters for a pressureless perfect fluid ($w_{m}=0$), a radiation fluid
($w=\frac{1}{3}$), and a stiff fluid ($w_{m}=1$) for $\lambda=-3$, $\lambda
=1$, and $\lambda=3$, and $\alpha>0$ (equivalent to $z>0$). }\label{FigA11}%
\end{figure}

Evaluating at the fixed points of \eqref{COSMOEVOL2}, we have found the
cosmological solutions:
\end{subequations}
\begin{align}
&  (\Omega_{\phi}, \alpha, \beta, \lambda)=\left(  1,\frac{1}{3}%
,1,\hat{\lambda}\right)  , H=\frac{1}{3 t}, q=2.\\
&  F^{+}(\hat{\lambda}): \left(  \frac{3 \left(  w_{m}+3\right)  }{2
\hat{\lambda}},\frac{2}{3 \left(  w_{m}+1\right)  },\frac{3 \left(
w_{m}+1\right)  }{\hat{\lambda}},\hat{\lambda} \right)  , H=\frac{2}{3 \left(
w_{m}+1\right)  }t^{-1}, q=\frac{1}{2} \left(  3w_{m}+1\right)  .\\
&  G^{+}(\hat{\lambda}): \left(  1,\frac{1}{\hat{\lambda}-3},2-\frac{6}%
{\hat{\lambda}},\hat{\lambda}\right)  , H=\frac{1}{\hat{\lambda}-3}t^{-1},
q=\hat{\lambda}-4.
\end{align}
where we used the notation $\hat{\lambda}=h^{(-1)}(0)$. Using the above
normalization, the result is the ``scaling away'' of the effects of the
overall expansion. However, in order to relate the analysis to observations,
the equations that determine the evolution of $H$, and clock time have to
brought into play \cite{WE}. The equations \eqref{COSMOEVOL2} can be written
as
\begin{subequations}
\label{COSMOEVOL3}%
\begin{align}
&  \frac{d{\Omega_{\phi}}}{d \ln t}=\alpha(\Omega_{\phi} -1) (2 q-3
w_{m}-1),\\
&  \frac{d{\alpha}}{d \ln t}=-\alpha(\alpha+\alpha q-1),\\
&  \frac{d{\beta}}{d \ln t}=\alpha(q (\beta-2)-3 (w_{m}+1) \Omega_{\phi}+3
w_{m}+4 \beta+1),\\
&  \frac{d{\lambda}}{d \ln t}=-\alpha\beta h(\lambda),
\end{align}
where
\end{subequations}
\begin{equation}
q=2+\lambda(\beta-\Omega_{\phi}).
\end{equation}
Since we have assumed $0\leq t<\infty$, then $-\infty<\ln t <+\infty$ is a
good time parameter for the dynamical system. The coordinates $(\Omega_{\phi},
\alpha, \beta, \lambda)$ of the fixed points of \eqref{COSMOEVOL2} can be
generically written as follows:

\begin{enumerate}
\item $\left(  \Omega_{\phi},0,\frac{q-2}{\lambda}+\Omega_{\phi}%
,\lambda\right)  $, eigenvalues $\left\{  0,0,0,\frac{\lambda-3 \lambda w_{m}
\left(  \Omega_{\phi}-1\right)  +\lambda(q+1) \Omega_{\phi}+q (-2
\lambda+q+2)-8}{\lambda}\right\}  $.

\item $\left(  1,0,\frac{q+\lambda-2}{\lambda},\lambda\right)  $, eigenvalues:
$0,0,0,\frac{(q-2) (-\lambda+q+4)}{\lambda} $.

\item $\left(  \Omega_{\phi},0,\frac{q-2}{\hat{\lambda}}+\Omega_{\phi}%
,\hat{\lambda}\right)  $, eigenvalues: $0,0,0,\frac{\hat{\lambda} \left(
\Omega_{\phi} \left(  -3 w_{m}+q+1\right)  +3 w_{m}-2 q+1\right)  +(q-2)
(q+4)}{\hat{\lambda}} $.

\item $\left(  \Omega_{\phi},0,\frac{2 q+3 w_{m} \left(  \Omega_{\phi
}-1\right)  +3 \Omega_{\phi}-1}{q+4}, -\frac{(q-2) (q+4)}{q \left(
\Omega_{\phi}-2\right)  -3 w_{m} \left(  \Omega_{\phi}-1\right)  +\Omega
_{\phi}+1}\right)  $, eigenvalues: $0,0,0,0 $.

\item $\left(  \frac{2-q}{\lambda},0,0,\lambda\right)  $, eigenvalues:
$0,0,0,\frac{\lambda+3 w_{m} (\lambda+q-2)+(3-2 \lambda) q-6}{\lambda} $.

\item $(1,0,0,2-q)$, eigenvalues:$0, 0, 0, -2 (1 + q) $.

\item $\left(  1,0,\frac{2 (q+1)}{q+4},q+4\right)  $, eigenvalues: $0,0,0,0 $.

\item $\left(  1,0,\frac{q-2}{\hat{\lambda}}+1,\hat{\lambda}\right)  $,
eigenvalues: $0,0,0,\frac{(q-2) \left(  -\hat{\lambda}+q+4\right)  }%
{\hat{\lambda}} $.

\item $\left(  \frac{-q^{2}-2 q+\hat{\lambda} \left(  2 q-3 w_{m}-1\right)
+8}{\hat{\lambda} \left(  q-3 w_{m}+1\right)  },0,\frac{\hat{\lambda} \left(
2 q-3 w_{m}-1\right)  -3 (q-2) \left(  w_{m}+1\right)  }{\hat{\lambda} \left(
q-3 w_{m}+1\right)  },\hat{\lambda}\right)  $, eigenvalues: $0,0,0,0 $.

\item $\left(  1-\frac{2 (q+1)}{3 \left(  w_{m}+1\right)  },0,0,\frac{3 (q-2)
\left(  w_{m}+1\right)  }{2 q-3 w_{m}-1}\right)  $, eigenvalues $0,0,0,0 $.

\item $\left(  1,\frac{1}{q+1},\frac{2 (q+1)}{q+4},q+4\right)  , \;
\text{with}\; h(4 + q)=0$, eigenvalues:
\begin{align*}
&  -\frac{q+\sqrt{24-q (q (4 q (q+1)-13)-28)}+4}{2 (q+1)^{2}},-\frac
{q-\sqrt{24-q (q (4 q (q+1)-13)-28)}+4}{2 (q+1)^{2}},\\
&  \frac{-3 w_{m}+2 q-1}{q+1},-\frac{2 h^{\prime}(q+4)}{q+4} .
\end{align*}

\end{enumerate}

For the exponential potential $h\equiv0$ and $\lambda$ becomes constant. Thus,
the system is reduced to one dimension, and the coordinates $(\Omega_{\phi
},\alpha,\beta)$ of the fixed points can be obtained explicitly as $\left(
\Omega_{\phi},0,\beta\right)  $, $\left(  1,\frac{1}{3},1\right)  $, $\left(
\frac{3\left(  w_{m}+3\right)  }{2\lambda},\frac{2}{3\left(  w_{m}+1\right)
},\frac{3\left(  w_{m}+1\right)  }{\lambda}\right)  $, $\left(  1,\frac
{1}{\lambda-3},2-\frac{6}{\lambda}\right)  $ (extensively studied in Appendix
\ref{appendixA}). In the Figure \ref{FigA11} is presented the evolution of the
system \eqref{COSMOEVOL2} for the exponential potential for some choices of
the parameters for a pressureless perfect fluid ($w_{m}=0$), a radiation fluid
($w=\frac{1}{3}$), and a stiff fluid ($w_{m}=1$) for $\lambda=-3$, $\lambda
=1$, and $\lambda=3$, and $\alpha>0$ (equivalent to $z>0$). In this case
observe that the points $D^{+}$ and $E^{+}$ both satisfy $\Omega_{\phi
}=1,\beta=1$, this is the first indication that the variables $\Omega_{\phi
},\beta$ are degenerated as phase space variables, but the diagram entails
relevant physical information about the cosmological observables. The case
$\lambda=6,w_{m}=1$ is not presented in this diagram (see at the figure
\eqref{fig:Fig2} the corresponding phase space plane $(x,y),z=+1$) since all
the points coalesce in one point which means that the diagram is highly
degenerated in these variables. For the choices $\lambda=6,w_{m}=0$,
$\lambda=6,w_{m}=\frac{1}{3}$ two points are degenerated and a third one is
close to them, so the dynamics on the plane $(\Omega_{\phi},\beta)$ is
obscure. All together, reinforces the idea that our variables $(x,y,z)$ are
more suitable for the description of the dynamics. For the other cases beyond
the exponential case, the plots in the plane $(\Omega_{\phi},\beta)$ resembles
many features of the exponential one, we do not present them by space.

From the Appendix \ref{appendixB} we extract that the generic solutions
includes: static solutions; static stiff solutions; decelerated contracting
stiff solutions; decelerated expanding stiff solutions; a line of de-Sitter
solutions; contracting accelerated de-Sitter solution; expanding accelerated
de-Sitter solution; ideal gas contracting scaling solutions; ideal gas
expanding scaling solutions; contracting scalar field dominated solution; and
expanding scalar field dominated solution. Some of these configurations
corresponds to values of $\lambda$ satisfying $h(\lambda)=0$. As shown, the
model at hand resembles a rich cosmological behavior, since it admits the
standard cosmological solutions and additionally it admits static solutions
and both expanding and contracting solutions. All these solutions have been
correlated with the fixed points of the system \eqref{cosmevol}.

\section{Conclusions}

\label{concl}

The determination of analytical solutions is essential in all areas of
physics. Concerning the gravitational theories, because of the nonlinearity of
the field equations, solutions which include all the free parameters are
difficult to be found, and for that, various methods from the analysis of
nonlinear differential equations and dynamical systems have been applied.

In this article we choose to work with the Cartan formalism and apply the
context of Cartan symmetries for the study of Liouville integrable systems in
a gravitational theory. In our model we considered that the universe is
isotropic and homogeneous where a scalar field, which attributes the degrees
of freedom of a higher-order modified teleparallel theory, is assumed to
describe the dark energy which drives the acceleration of the universe.

From the different kind of Cartan symmetries, which the field equations can
admit, we considered those symmetries which are linear in the first
derivatives. The field equations are rational in the momentum/first
derivatives, therefore, conservation laws rational in the momentum are
favored. Moreover, we saw that the systems which admit Cartan symmetries
linear in the momentum include a big range of possible dynamical systems
including those which are invariant under point transformations.

Our analysis provided four families of potentials where there exists a
dependence on the parameters of the potentials with the constant equation of
state parameter for the matter source. This kind of dependence has been
observed before in other cosmological models \cite{lie5,gian}. For those
models the Cartan symmetries and the corresponding conservation laws were
determined while the solution of the Hamilton-Jacobi equation has been
derived. Furthermore, the field equations have been reduced to a system of two
first-order differential equations which is the analytical solution.
Closed-form solutions, and some exact solutions have been derived, for
specific values of the integration constants, while the behaviour of the
solution at late times was studied.

In particular we found that the noncanonical scalar field provides a
cosmological constant term, stiff fluid components as the quintessence field
but also dark matter components can be introduced like the unified dark model
\cite{aasen}. Last but not least, we saw that scale factors which describe the
inflation era can be determined.

Furthermore, from a closed-form solution that we derived, we wrote the Hubble
function in terms of the scale factor and we compared that toy model with the
Supernova data. We saw that this model fits the standard candles in a similar
way with that of $\Lambda$-cosmology, since both cosmologies contain the same
number of free parameters.

However in order to perform a global study for the evolution of that theory we
performed an extendent critical point analysis by using coordinates different
from those of the Hubble-normalization, such an analysis is important because
provide results also for non-integrable models. Indeed, the Hubble function
$H\left(  t\right)  $ can cross the value $H\left(  t\right)  =0$, from
negative to positive values, or vice-versa, since $\rho_{\phi}$ can be
negative due the friction term $3H\dot{\phi}$. This implies that the
Hubble-normalization procedure allows only to describe just a patch of the
phase space. In particular, we use more proper phase-space variables first
introduced in \cite{Giacomini:2017yuk}.

To analyze the fixed point for arbitrary potentials, we have used the method
called in our notation $h$-devisers, which allows us to perform the whole
analysis for a wide range of potentials
\cite{Fang:2008fw,Matos:2009hf,Leyva:2009zz,UrenaLopez:2011ur,Copeland:2009be,Farajollahi:2011ym,
Xiao:2011nh,Escobar:2011cz,Escobar:2012cq,
Escobar:2013js,delCampo:2013vka,Fadragas:2013ina}. Using this method, we have
studied the exponential potential and non- exponential potentials for which
$h(\lambda)$ can be written in an explicit form, e.g, $V(\phi)=V_{0}%
e^{-\sigma\phi}+V_{1},\sigma\neq0$, $h\equiv-\lambda(\lambda-\sigma)$;
$V\left(  \phi\right)  =\frac{(\mu\phi)^{k}}{k}$ with $h\equiv-\frac
{\lambda^{2}}{k}$; $V(\phi)=V_{0}(\cosh(\xi\phi)-1),\xi\neq0$, with
$h\equiv-\frac{1}{2}\left(  \lambda^{2}-\xi^{2}\right)  $; $V_{B}\left(
\phi\right)  =V_{1}e^{-3(w_{m}+1)\phi}+V_{2}e^{-6w_{m}\phi}$, with
$h_{B}\equiv-\left(  \lambda-3(w_{m}+1)\right)  (\lambda-6w_{m})~$, and
$V_{C}\left(  \phi\right)  =V_{1}e^{-3\left(  1+w_{m}\right)  \phi}%
+V_{2}e^{-\frac{3}{2}\left(  3+w_{m}\right)  }$, with $h_{C}\equiv-\frac{1}%
{2}\left(  \lambda-3(w_{m}+1)\right)  \left(  2\lambda-3(3+w_{m})\right)  $.
The last two models were introduced by the Cartan symmetries in Section
\ref{cartan2}. We have found that there are generic solutions: static
solutions; static stiff solutions; decelerated contracting stiff solutions;
decelerated expanding stiff solutions; a line of de-Sitter solutions;
contracting accelerated de-Sitter solution; expanding accelerated de-Sitter
solution; ideal gas contracting scaling solutions; ideal gas expanding scaling
solutions; contracting scalar field dominated solution; and expanding scalar
field dominated solution. Some of these configurations corresponds to values
of $\lambda$ satisfying $h(\lambda)=0$.
As showed, the model at hand resembles a rich cosmological behavior, since it
admits the standard cosmological solutions and additionally it admits static
solutions and both expanding and contracting solutions. All these solutions
were correlated with the fixed points of the system \eqref{cosmevol}. Finally,
we have investigated the evolution of the observables, the so called age
parameter $\alpha=tH$, the deceleration parameter $q$, and the fractional
energy of scalar field and Hubble-normalized kinetic term in a phase space.
Imposing observational constraints on the current values of $\alpha_{0}%
=\alpha(\mathbf{y}_{0})$, and the matter parameter $\Omega_{0}=1-\Omega_{\phi
}(\mathbf{y}_{0})$, it is restricted the location of the present state of the
universe, $\mathbf{y}_{0}$, in state space.

This work extents our research program on the geometric selection rules in
gravitational theories and on the determination of analytical solutions as
also on the role of symmetries in the evolution of the universe.

\begin{acknowledgments}
AP acknowledges the financial support of FONDECYT grant no. 3160121 and thanks
the University of Athens for the hospitality provided while this work carried
out. GL thanks to Department of Mathematics at Universidad Catolica del Norte
for warm hospitality and financial support.
\end{acknowledgments}

\appendix

\section{Hubble-normalization}

\label{appendixA}

For the completness of our analysis and compare our results with that of
\cite{an1}. We present the fixed point analysis for the field equations by
using the Hubble-normalization, that is, by defining the new variables to be
\begin{equation}
\beta=\frac{\dot{\phi}}{H},\quad\chi=\frac{V(\phi)}{6H^{2}},
\end{equation}
related through the constraint equations
\begin{equation}
\Omega_{m}+\beta+\chi=1,
\end{equation}
and introducing the new time derivative
\[
\tilde{f}:=\frac{\dot{f}}{|H|}=\frac{\dot{f}}{\left\vert z\right\vert }%
,~z\neq0.
\]

This gives the lower dimensional dynamical system
\begin{subequations}
\label{FRW-H-normalization}%
\begin{align}
&  \tilde{\beta}=-\epsilon\left(  \chi\left(  -2\lambda+3w_{m}+\lambda
\beta+3\right)  +3(w_{m}-1)\left(  \beta-1\right)  \right)  ,\\
&  \tilde{\chi}=-\epsilon\chi\left(  \lambda\beta+2\lambda\chi-6\right)  ,\\
&  \tilde{\lambda}=-\epsilon\beta h(\lambda).
\end{align}
where $\epsilon$ is the sign of $H$.

This system is not well defined when $z$ changes sign; however, it can
describe the regions of the phase space $H<0$ or $H>0$. Notice that the fixed
points corresponding to contracting universes ($H<0$) will have the reverse
dynamical behavior of the analogous points with $H>0$, such that we can
restrict our attention to expanding models in the cosmological applications.

We discuss briefly on the stability of the fixed points of
\eqref{FRW-H-normalization}. In the notation the subscript $\epsilon=\pm1$
corresponds to the sign of $z$, that gives if the model corresponds to
expansion ($\epsilon=+1$) or to contraction ($\epsilon=-1$) as in
\cite{Campos:2001pa} (see references therein). \footnote{We don't use
superscripts to do not mix with the notation used in sections \ref{Sec:3.1}
and \ref{Sec:3.2}, but the fixed points are closely related.} For the choice
$\epsilon=+1$ are recovered all the results presented in \cite{an1}.

For the exponential potential (for which $\lambda$ is constant) we have the
fixed points
\end{subequations}
\begin{enumerate}
\item $D_{\epsilon}:(\beta,\chi)=(0,1)$. Exists for $\lambda=3$. The
eigenvalues are $-3(w_{m}+1)\epsilon, -3\epsilon$.

\begin{enumerate}
\item The fixed points $D_{\epsilon}$ are nonhyperbolic for $w_{m}=-1$.

\item $D_{+}$ (respectively, $D_{-}$) is stable (respectively, unstable) for
$w_{m}>-1$.

\item They are saddles otherwise.
\end{enumerate}

\item $E_{\epsilon}:(\beta,\chi)=(1,0)$. The eigenvalues are $-3(w_{m}%
-1)\epsilon,(6-\lambda)\epsilon$.

\begin{enumerate}
\item The points are nonhyperbolic for $w_{m}=1$ or $\lambda=6$.

\item The fixed point $E_{+}$ (respectively, $E_{-}$) is a source
(respectively, a sink), for $w_{m}<1,\lambda<6$.

\item They are saddles otherwise.
\end{enumerate}

\item $F_{\epsilon}:(\beta,\chi)=\left(  \frac{3(w_{m}+1)}{\lambda}%
,-\frac{3(w_{m}-1)}{2\lambda}\right)  $. The eigenvalues are \newline$\frac
{1}{4}\epsilon\left(  3w_{m}-3+\sqrt{3}\sqrt{(1-w_{m})(-16\lambda+21w_{m}%
+75)}\right)  $,\newline$\frac{1}{4}\epsilon\left(  3w_{m}-3-\sqrt{3}%
\sqrt{(1-w_{m})(-16\lambda+21w_{m}+75)}\right)  $.

\begin{enumerate}
\item The points are nonhyperbolic for $\lambda=\frac{3(w_{m}+3)}{2}$, or
$w_{m}=1$.

\item The fixed point $F_{+}$ (respectively, $F_{-}$) is a sink (respectively,
a source), for $w_{m}<1,\lambda>\frac{3}{2}(w_{m}+3)$.

\item They are saddles otherwise.
\end{enumerate}

\item $G_{\epsilon}:(\beta,\chi)=\left(  2-\frac{6}{\lambda},\frac{6}{\lambda
}-1\right)  $. The eigenvalues are $(\lambda-6)\epsilon,\epsilon
(2\lambda-3w_{m}-9)$.

\begin{enumerate}
\item The points are nonhyperbolic for either $\lambda=6$ or $\lambda
=\frac{3(w_{m}+3)}{2}$.

\item The fixed point $G_{+}$ (respectively, $G_{-}$) is a sink (respectively,
a source) for $w_{m}\leq1,\lambda<\frac{3}{2}(w_{m}+3)$.

\item The fixed point $G_{+}$ (respectively, $G_{-}$) is a source
(respectively, a sink) for $w_{m}\leq1,\lambda>6$.

\item They are saddle otherwise.
\end{enumerate}
\end{enumerate}

For the arbitrary potentials we obtain the fixed points

\begin{enumerate}
\item $D_{\epsilon}:(\beta,\chi,\lambda)=(0,1,3)$. Always exists. The
eigenvalues are $-3(w_{m}+1)\epsilon$, \newline$\frac{1}{2}\left(
-\sqrt{9-8h(3)}-3\right)  \epsilon,\frac{1}{2}\left(  \sqrt{9-8h(3)}-3\right)
\epsilon$.

\begin{enumerate}
\item The fixed points $D_{\epsilon}$ are nonhyperbolic for $w_{m}=-1$ or
$h(3)=0$.

\item $D_{+}$ (respectively, $D_{-}$) is stable (respectively, unstable) for
$w_{m}>-1,h(3)>0$.

\item They are saddles otherwise.
\end{enumerate}

\item $E_{\epsilon}(\hat{\lambda}):(\beta,\chi,\lambda)=(1,0,\hat{\lambda})$,
such that $h(\hat{\lambda})=0$. The eigenvalues are \newline$-3(w_{m}%
-1)\epsilon,-\epsilon\left(  \hat{\lambda}-6\right)  ,-\epsilon h^{\prime
}\left(  \hat{\lambda}\right)  $.

\begin{enumerate}
\item The points are nonhyperbolic for $w_{m}=1$ or $\hat{\lambda}=6$ or
$h^{\prime}\left(  \hat{\lambda}\right)  =0$.

\item The fixed points $E_{+}(\hat{\lambda})$ (respectively, $E_{-}%
(\hat{\lambda})$) are sources (respectively, a sink), for $w_{m}%
<1,\hat{\lambda}<6,h^{\prime}\left(  \hat{\lambda}\right)  <0$.

\item They are saddles otherwise.
\end{enumerate}

\item $F_{\epsilon}(\hat{\lambda}):(\beta,\chi,\lambda)=\left(  \frac
{3(w_{m}+1)}{\hat{\lambda}},-\frac{3(w_{m}-1)}{2\hat{\lambda}},\hat{\lambda
}\right)  $, such that $h(\hat{\lambda})=0$. The eigenvalues are $\frac{1}%
{4}\epsilon\left(  3w_{m}-3-\sqrt{3}\sqrt{(1-w_{m})\left(  -16\hat{\lambda
}+21w_{m}+75\right)  }\right)  $, \newline$\frac{1}{4}\epsilon\left(
3w_{m}-3+\sqrt{3}\sqrt{(1-w_{m})\left(  -16\hat{\lambda}+21w_{m}+75\right)
}\right)  $, $-\frac{3(w_{m}+1)\epsilon h^{\prime}\left(  \hat{\lambda
}\right)  }{\hat{\lambda}}$.

\begin{enumerate}
\item The points are nonhyperbolic for either $w_{m}=-1$, or $\hat{\lambda
}=\frac{3(w_{m}+3)}{2}$, or $w_{m}=1$, or $h^{\prime}(\hat{\lambda})=0$

\item The fixed points $F_{+}(\hat{\lambda})$ (respectively, $F_{-}%
(\hat{\lambda})$) are sinks (respectively, sources) for

\begin{enumerate}
\item $h^{\prime}\left(  \hat{\lambda}\right)  >0,-1<w_{m}<1,\hat{\lambda
}>\frac{3}{2}(w_{m}+3)$
\end{enumerate}

\item They are saddle otherwise.
\end{enumerate}

If we restrict the equation of state on the range $-1\leq w_{m}\leq1$, just
the cases (a), (b)-(iv) and (d) apply.

\item $G_{\epsilon}(\hat{\lambda}):(\beta,\chi)=\left(  2-\frac{6}%
{\hat{\lambda}},\frac{6}{\hat{\lambda}}-1,\hat{\lambda}\right)  $. The
eigenvalues are \newline$\left(  \hat{\lambda}-6\right)  \epsilon
,\epsilon\left(  2\hat{\lambda}-3(w_{m}+3)\right)  ,-\frac{2\left(
\hat{\lambda}-3\right)  \epsilon h^{\prime}\left(  \hat{\lambda}\right)
}{\hat{\lambda}}$.

\begin{enumerate}
\item The points are nonhyperbolic for either $\hat{\lambda}=6$ or
$\hat{\lambda}=3$ or $\hat{\lambda}=\frac{3(w_{m}+3)}{2}$ or $h^{\prime}%
(\hat{\lambda})=0$.

\item The fixed points $G_{+}(\hat{\lambda})$ (respectively, $G_{-}%
(\hat{\lambda})$) are sinks (respectively, sources) for

\begin{enumerate}
\item $\hat{\lambda}<0,h^{\prime}\left(  \hat{\lambda}\right)  >0$, or

\item $0<\hat{\lambda}<3,h^{\prime}\left(  \hat{\lambda}\right)  <0$, or

\item $3<\hat{\lambda}<6,\frac{1}{3}\left(  2\hat{\lambda}-9\right)
<w_{m}\leq1,h^{\prime}\left(  \hat{\lambda}\right)  >0$.
\end{enumerate}

\item The fixed points $G_{+}(\hat{\lambda})$ (respectively, $G_{-}%
(\hat{\lambda})$) are sources (respectively, sinks) for $\hat{\lambda
}>6,h^{\prime}\left(  \hat{\lambda}\right)  <0$.

\item they are saddle otherwise.
\end{enumerate}
\end{enumerate}

\section{Fixed points of the system \eqref{COSMOEVOL3}}

\label{appendixB}

The coordinates $(\Omega_{\phi}, \alpha, \beta, \lambda, q)$ of the fixed
points of the system \eqref{COSMOEVOL3} are:

\begin{enumerate}
\item $\left(  \Omega_{\phi},0,\beta,\lambda,\beta\lambda-\Omega_{\phi}
\lambda+2\right)  .$

\item $\left(  \Omega_{\phi},0,\frac{3 \left(  w_{m}-1\right)  }{2 \lambda
}+\Omega_{\phi},\lambda,\frac{1}{2} \left(  3 w_{m}+1\right)  \right)  $.

\item $(1,0,\beta,\lambda,(\beta-1) \lambda+2)$.

\item $\left(  \Omega_{\phi},0,0,\lambda,2-\lambda\Omega_{\phi}\right)  $.

\item $\left(  \Omega_{\phi},0,\beta,\frac{3 \left(  -2 \beta+w_{m} \left(
\Omega_{\phi}-1\right)  +\Omega_{\phi}+1\right)  }{(\beta-2) \left(
\beta-\Omega_{\phi}\right)  },\frac{-4 \beta+3 w_{m} \left(  \Omega_{\phi
}-1\right)  +3 \Omega_{\phi}-1}{\beta-2}\right)  $.

\item $\left(  \Omega_{\phi},0,\beta,\hat{\lambda},\hat{\lambda} \left(
\beta-\Omega_{\phi}\right)  +2\right)  $.

\item $\left(  \Omega_{\phi},0,\frac{3 \left(  w_{m}-1\right)  }{2
\hat{\lambda}}+\Omega_{\phi},\hat{\lambda},\frac{1}{2} \left(  3
w_{m}+1\right)  \right)  $.

\item $\left(  -\frac{3 \left(  w_{m}-1\right)  }{2 \lambda},0,0,\lambda
,\frac{1}{2} \left(  3 w_{m}+1\right)  \right)  $.

\item $\left(  1,0,\beta,-\frac{6}{\beta-2},-4-\frac{6}{\beta-2}\right)  $.

\item $(1,0,0,\lambda,2-\lambda)$.

\item $\left(  \frac{w_{m}+3}{w_{m}+1},0,2,\lambda,\frac{\lambda\left(
w_{m}-1\right)  }{w_{m}+1}+2\right)  $.

\item $\left(  \Omega_{\phi},0,0,\frac{3 \left(  w_{m} \left(  \Omega_{\phi
}-1\right)  +\Omega_{\phi}+1\right)  }{2 \Omega_{\phi}},\frac{1}{2} \left(  3
w_{m}-3 \left(  w_{m}+1\right)  \Omega_{\phi}+1\right)  \right)  $.

\item $\left(  1,0,\beta,\hat{\lambda},(\beta-1) \hat{\lambda}+2\right)  $.

\item $\left(  \frac{(\beta-2) \hat{\lambda} \beta+6 \beta+3 w_{m}-3}%
{(\beta-2) \hat{\lambda}+3 w_{m}+3},0,\beta,\hat{\lambda},\frac{6 \left(
w_{m}+1\right)  +\hat{\lambda} \left(  -\beta+3 (\beta-1) w_{m}-1\right)
}{(\beta-2) \hat{\lambda}+3 w_{m}+3}\right)  $.

\item $\left(  \frac{\beta\left(  w_{m}+3\right)  }{2 \left(  w_{m}+1\right)
},0,\beta,\frac{3 \left(  w_{m}+1\right)  }{\beta},\frac{1}{2} \left(  3
w_{m}+1\right)  \right)  $.

\item $\left(  1,0,0,3,-1\right)  $.

\item $\left(  1,\frac{1}{3},1,\hat{\lambda},2\right)  $.

\item $(1,0,1,\lambda,2)$.

\item $\left(  \frac{3 \left(  w_{m}+3\right)  }{2 \hat{\lambda}},\frac{2}{3
\left(  w_{m}+1\right)  },\frac{3 \left(  w_{m}+1\right)  }{\hat{\lambda}%
},\hat{\lambda},\frac{1}{2} \left(  3 w_{m}+1\right)  \right)  $.

\item $\left(  \frac{3 \left(  w_{m}+3\right)  }{2 \hat{\lambda}},0,\frac{3
\left(  w_{m}+1\right)  }{\hat{\lambda}},\hat{\lambda},\frac{1}{2} \left(  3
w_{m}+1\right)  \right)  $.

\item $\left(  1,\frac{1}{\hat{\lambda}-3},2-\frac{6}{\hat{\lambda}}%
,\hat{\lambda},\hat{\lambda}-4\right)  $.

\item $\left(  1,0,2-\frac{6}{\hat{\lambda}},\hat{\lambda},\hat{\lambda
}-4\right)  $.
\end{enumerate}

\end{document}